\documentclass[aps,prx,10pt,twocolumn,floatfix,superscriptaddress,showpacs]{revtex4-2}

\usepackage{amsmath, amssymb}
\usepackage{enumitem}
\usepackage{mathtools}
\usepackage{lipsum}
\usepackage{times}
\usepackage{graphicx}
\usepackage{wasysym}
\usepackage{bm}
\usepackage{hyperref}
\usepackage{xspace}
\usepackage{siunitx}
\usepackage{soul}
\usepackage{braket}
\usepackage[normalem]{ulem}
\usepackage{orcidlink}
\usepackage[T1]{fontenc}

\usepackage{dcolumn}
\usepackage{bm}
\usepackage{mathrsfs}
\usepackage{mathtools}
\usepackage{physics}
\usepackage{upgreek}

\definecolor{myColor}{rgb}{0,0,0.8}
\definecolor{myciteColor}{rgb}{0.39,0.7,0.89}
\hypersetup{colorlinks=true, linkcolor=myColor, filecolor=myColor, urlcolor=myColor,citecolor=myColor, urlcolor=myColor}
\graphicspath{{./Figures/}}

\DeclareSIUnit{\nK}{\nano\kelvin}
\DeclareSIUnit{\aB}{\emph{a}_0}
\DeclareSIUnit{\G}{G}

\newcommand{\kB}{k_{\text{B}}}

\newcommand{\lambdac}{\lambda_\mathrm{c}}
\newcommand{\omegac}{\omega_\mathrm{c}}
\newcommand{\Nc}{N_\mathrm{c}}
\newcommand{\GDown}{\Gamma_{\downarrow}}
\newcommand{\GUp}{\Gamma_{\uparrow}}
\newcommand{\Babs}{B_\mathrm{abs}}
\newcommand{\Bem}{B_\mathrm{em}}
\newcommand{\Mg}{M_\mathrm{g}}
\newcommand{\Me}{M_\mathrm{e}}
\newcommand{\myoverbar}[1]{\mkern 0.5mu\overline{\mkern-3mu#1\mkern-1.5mu}\mkern -1.5mu} 
\newcommand{\mytilde}[1]{\mkern 1.0mu\widetilde{\mkern-1mu#1\mkern-1mu}\mkern -1.5mu} 
\newcommand{\Mgb}{\myoverbar{M}_\mathrm{g}}
\newcommand{\Meb}{\myoverbar{M}_\mathrm{e}}
\newcommand{\Mgt}{\mytilde{M}_\mathrm{g}}
\newcommand{\Met}{\mytilde{M}_\mathrm{e}}
\newcommand{\nb}{\bar{n}}
\newcommand{\nt}{\widetilde{n}}

\newcommand\ddfrac[2]{\frac{\displaystyle #1}{\displaystyle #2}}

\begin{document}
\title{Time-periodic driving of a bath-coupled open quantum gas of light}

\author{Andris Erglis\,\orcidlink{0000-0002-6842-4876}}
\altaffiliation[Present address: ]{Università degli Studi di Palermo, Dipartimento di Fisica e Chimica-Emilio Segrè (UNIPA), Via Archirafi 36, 90123 Palermo, Italy.}
\affiliation{Physikalisches Institut, Albert-Ludwigs-Universität Freiburg, Hermann-Herder-Straße 3, 79104, Freiburg, Germany}

\author{Alexander Sazhin\,\orcidlink{0009-0009-1915-8696}}
\affiliation{Institut f\"ur Angewandte Physik, Universit\"at Bonn, Wegelerstr. 8, 53115 Bonn, Germany}

\author{Frank Vewinger\,\orcidlink{0000-0001-7818-2981}}
\affiliation{Institut f\"ur Angewandte Physik, Universit\"at Bonn, Wegelerstr. 8, 53115 Bonn, Germany}

\author{Martin Weitz\,\orcidlink{0000-0002-4236-318X}}
\affiliation{Institut f\"ur Angewandte Physik, Universit\"at Bonn, Wegelerstr. 8, 53115 Bonn, Germany}

\author{Stefan Yoshi Buhmann}
\affiliation{Institut für Physik, University of Kassel, Heinrich-Plett-Str. 40, 34132 Kassel, Germany}

\author{Julian Schmitt\,\orcidlink{0000-0002-0002-3777}}
\email[Corresponding author: ]{julian.schmitt@kip.uni-heidelberg.de}
\affiliation{Institut f\"ur Angewandte Physik, Universit\"at Bonn, Wegelerstr. 8, 53115 Bonn, Germany}
\affiliation{Kirchhoff-Institut f{\"u}r Physik, Universit{\"a}t Heidelberg, Im Neuenheimer Feld 225a, 69120 Heidelberg, Germany}

\date{\today}

\begin{abstract}
We study the frequency-resolved density response of a photon Bose--Einstein condensate coupled to a bath of dye molecules by time-periodic driving. By monitoring the photon number dynamics for different drive frequencies, we obtain the spectral response of the condensate in a phase-sensitive way. We find that as the photon number increases, the response of the coupled condensate-bath system transitions from overdamped to resonant behavior, indicating a transition from closed to open system dynamics. Our spectroscopy method paves the way for studies of collective excitations in complex driven-dissipative systems.
\end{abstract}



\maketitle

{\itshape{Introduction.---}} Temporally controlling many-body systems on the timescale of their microscopic dynamics plays an important role in many physical situations, from ultrafast manipulation of solid-state quantum materials~\cite{Oka:2019} to quantum simulation with ultracold atomic gases~\cite{Eckardt:2017}. A well-known example of coherent control is Floquet engineering, which relies on the time-periodic modulation of a Hamiltonian to tailor the quasi-energy band structure~\cite{Holthaus:2015,Castro:2022}, similar to Bloch bands in spatially periodic potentials. In coherent systems with negligible dissipation, such methods have enabled, for example, the realization of Floquet-topological insulators in arrays of photonic waveguides~\cite{Rechtsman:2013} or ultracold fermions in optical lattices~\cite{Jotzu:2014}, as well as of interacting Floquet polaritons~\cite{Clark:2019}.

In dissipative systems, the implications of periodic driving are much less explored. Understanding the interplay between driving and dissipation is a fundamental problem in nonequilibrium statistical physics that has recently attracted growing theoretical interest~\cite{Ikeda:2020,Sato:2020,Ochoa:2024}, due to improved experimental techniques to control dissipation~\cite{Barreiro:2011,Barontini:2013,Tomita:2017,Wetter:2023}, e.g., for non-Hermitian Floquet engineering~\cite{Fedorova:2020}. A new approach in this direction is based on quantum gases of light and matter, which other than atomic gases can be easily coupled to periodically-driven reservoirs. Recent work on exciton-polariton condensates has, for instance, demonstrated the injection of quantized vortices by spatio-temporally stirring a laser that pumps the nonequilibrium system~\cite{Gnusov:2023,delValle:2023,delValle:2024}. Photon Bose--Einstein condensates inside material-filled optical microcavities, on the other hand, provide a suitable platform to study reservoir-induced driving also close to thermal equilibrium~\cite{Klaers:2010,Marelic:2015,Greveling:2018,Schofield:2024,Pieczarka:2024}. In this platform, a two-dimensional photon gas is coupled to dye molecules and thermalizes by absorption and emission processes~\cite{Schmitt:2015}. To drive the system, the chemical potential of the photon gas can be externally controlled in a time-dependent manner by optically pumping the molecules comprising the bath. Previous work has demonstrated the grand canonical coupling to a particle reservoir~\cite{Schmitt:2014} and verified a fluctuation--dissipation relation governing the particle number fluctuations~\cite{Oeztuerk:2023}. More recently, also the time-dependent control of the dye bath using a short pulse perturbation was successfully employed to probe the regression theorem in the optical quantum gas~\cite{Sazhin:2024,Bode:2024}.

In this Letter, we study the temporal control of an open quantum gas coupled to a bath by periodic driving. Our experiment utilizes a photon Bose--Einstein condensate (BEC) inside a dye-filled microcavity. In the weakly-coupled light--matter system, time-periodic modulation of the pump feeding the bath forces the photons to undergo periodic dynamics and thus leave their steady state. By monitoring the oscillations for different driving frequencies, we obtain the spectral amplitude and phase of the condensate response. For increasing photon numbers, our spectroscopy of the driven--dissipative quantum gas reveals a transition from a dissipative overdamped to an underdamped behavior, where a resonance in the response of the system appears. Physically, the resonance originates from weak losses and pumping (ns timescale), here being much slower than thermalization (ps timescale)~\cite{Kirton:2013,Schmitt:2015,Hesten:2018,Oeztuerk:2021,Sazhin:2024,Bode:2024}, which realize an effective restoring force of the BEC population back to its steady state. For large losses, this resonance phenomenon is
expected to cross over to the relaxation oscillations known in
laser physics. Our findings can be interpreted as a spectroscopic measurement of the elementary excitations of the open quantum gas, whose nature can be understood as the collective dynamics of weakly coupled photons and molecules, and therefore differentiates itself from those observed in cold atoms or polaritons, where the strongly coupled limit was applicable~\cite{Stamper-Kurn:1999b,Utsunomiya:2008}. Due to the interplay between the dissipative mechanisms of cavity loss and molecule decay, we predict a previously unseen cusp singularity in the phase diagram of the system.

\begin{figure}[b]
    \centering
   \includegraphics[width=1.0\columnwidth]{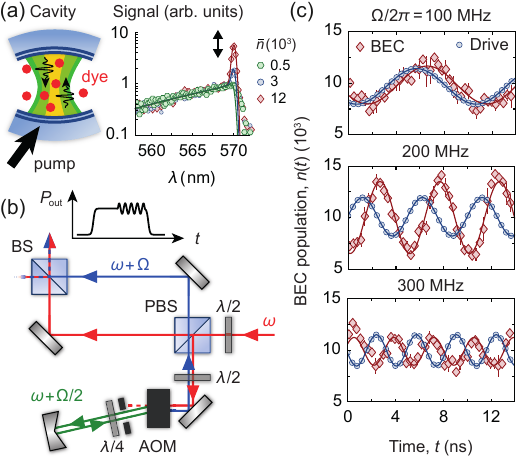} 
\caption{Experimental scheme. (a)~Dye-filled optical microcavity with photon BEC, non-resonantly injected by pumping the dye medium with a temporally-modulated laser beam. Right: Time-averaged photon gas spectra showing Bose--Einstein distributions at $T=\SI{300}{\kelvin}$ (lines). (b)~The time-periodic driving at frequency $\Omega$ is implemented by temporally modulating the pump power in a double-pass setup. Interference between laser fields at $\omega$ and $\omega + \Omega$ results in a low-amplitude beat note in the second half of the pump pulse where the AOM is switched on. (c) Exemplary time traces of the oscillating driving laser power (blue circles) and the response of the BEC population at $\nb=9000$ (red diamonds) for three modulation frequencies, along with fits (lines). Error bars in (c) denote statistical errors.}\label{fig:1}
\end{figure} 

{\itshape{Experimental method.---}} We prepare the photon BEC inside an optical microcavity formed by two mirrors with a $\SI{1}{\meter}$ radius of curvature spaced by $D_0\approx \SI{1.4}{\micro\meter}$, as shown in Fig.~\ref{fig:1}(a); for a detailed description, see Refs.~\cite{Klaers:2010,Schmitt:2018} and the SI~\cite{supplementalMaterial}. The cavity is filled with a liquid solution of Rhodamine 6G dye with refractive index $n_{\mathrm{r}}\approx 1.44$ and dye concentration $\SI{1}{\milli\mol\per\liter}$. In the short cavity, the free spectral range $\Delta\nu_\mathrm{FSR}=c/2n_\mathrm{r}D_0\approx\SI{74}{\tera\hertz}$ between adjacent longitudinal modes around the mode number $q=7$ is larger than the dye absorption and emission spectral bandwidth (of order $\kB T\approx 2\pi\hbar\times \SI{6.3}{\tera\hertz}$). Therefore, the longitudinal degree of freedom of the photons is thermodynamically frozen out, and the photon dynamics is restricted to the transverse cavity modes, which makes the system effectively two-dimensional. The ground state energy is determined by the cutoff wavelength $\lambdac=2n_{\mathrm{r}} D_0/q=\SI{570}{\nano\meter}$ and the mirror curvature realizes a harmonic trapping potential of frequency $\omega_{\mathrm{trap}}/2\pi\approx\SI{40}{\giga\hertz}$. To inject photons into the cavity, the dye medium is pumped with a laser beam near $\SI{532}{\nano\meter}$ wavelength chopped into $\SI{600}{\nano\second}$ long pulses every $\SI{20}{\milli\second}$ using acousto-optic modulators (AOMs)~\cite{supplementalMaterial}. The cavity photons thermalize by repeated absorption and emission processes with the dye medium, which fulfills the Kennard--Stepanov relation $\Bem/\Babs = \exp(-\hbar\Delta/\kB T)$ connecting the rates for emission and absorption $B_\mathrm{em,abs}$ of a photon by a dye molecule. Here, $T=\SI{300}{\kelvin}$ denotes the temperature and $\Delta=\omegac-\omega_{\mathrm{zpl}}$ the detuning between the cavity ground mode $\omegac/2\pi=c/\lambdac$ and the zero-phonon line $\omega_{\mathrm{zpl}}$ of the molecules~\cite{Schmitt:2018}; for the used $\hbar\Delta=-3.9\kB T$, one has $\Babs=\SI{492}{\hertz}$ and $\Bem=\SI{23.3}{\kilo\hertz}$~\cite{Schmitt:2024}. Above the critical photon number $N_\mathrm{c} = \pi^2/3 (\kB T)^2/(\hbar\omega_{\mathrm{trap}})^2 \approx 80000$, the gas enters the BEC phase and we observe a macroscopic occupation of the lowest-energy mode, along with an equilibrium distribution of excited states, see the right panel in Fig.~\ref{fig:1}(a).

To periodically drive the particle number in the photon BEC, we amplitude-modulate the pumping of the dye molecule bath. The oscillating pump power is realized by combining a double-pass setup with a Mach--Zehnder interferometer, as shown in Fig.~\ref{fig:1}(b). The incident laser beam of frequency $\omega$ is split by a polarizing beam splitter (PBS), where the reflected part twice passes an AOM operated at a variable radiofrequency $\Omega_\mathrm{RF}=\Omega/2$, with $\Omega_\mathrm{RF}/2\pi= \SI{30}{}$ to $\SI{160}{\mega\hertz}$. In each pass, the beam is diffracted to the first order, which in total shifts its frequency to $\omega+\Omega$. At the interferometer output, both beams are spatially overlapped by a nonpolarizing beam splitter (BS) to produce a beating signal modulating the pump rate $R(t)=\GUp + R_0\sin({\Omega t})$ of the optical microcavity~\cite{supplementalMaterial}. Here, $R_0$ is the driving amplitude of the pump. Small $R_0$, quintessential for weak driving in the regime of linear response, are realized by adjusting the ratio of reflection and transmission at the PBS with a $\lambda/2$-plate. In each experiment cycle, we determine the steady-state photon number that results from the pump rate $\GUp$ before the driving is started; for this, the modulation is switched on only in the second half of the pump pulse, see Fig.~\ref{fig:1}(b). The time-resolved photon number evolution in the condensate is monitored by recording the cavity emission on a photomultiplier~\cite{Sazhin:2024}. We also record the temporal evolution of the pump power, which allows us to determine the relative phase between the condensate response and the driving.

{\itshape{Theoretical model.---}} We model the photon BEC coupled to a periodically-driven bath of molecules in the weak-driving limit using rate equations that describe the coupling between $n$ photons, $\Mg$ ground, and $\Me$ excited state molecules:
\begin{equation}
\begin{aligned}\label{eq:rateequations}
\dot{n}= & -\Babs \Mg n+\Bem \Me (n+1) -\kappa n,\\
\dot{\Me}= & \Babs\Mg n -\Bem\Me (n+1)\!-\!\GDown \Me+R(t)\Mg. 
\end{aligned}
\end{equation}
Here, $\GDown$ denotes a molecule loss rate (e.g. from spontaneous emission into higher modes) and $\kappa$ the cavity photon loss rate. The bath driving is included by a time-dependent, oscillating pump rate $R(t)$ in the last term, which can be obtained from a semiclassical description of the photon--molecule interaction~\cite{Erglis:2022}. We make an ansatz for the photon and excited molecule dynamics around the steady-state values $\nb$ and $\Meb$, respectively, $n(t)=\nb+\nt(t)=\nb+|n_0(\Omega)|\sin(\Omega t+\phi)$ and $\Me(t)=\Meb+\Met(t)=\Meb+|M_{\mathrm{e}_0}(\Omega)|\sin(\Omega t+\phi_{M_{\mathrm{e}}})$, which oscillate at the drive frequency with variable phases. By inserting this into Eq.~\eqref{eq:rateequations} and neglecting higher-order terms $\sim\nt\Met$, we arrive at the linearized equation of motion of the periodically-driven photon condensate,
\begin{equation}\label{eq:drivenBECdiffEq}
   \ddot{\nt}+\Gamma \dot{\nt}+\Omega_0^2 \nt=F_0 \sin{\Omega t}. 
\end{equation}
The free oscillation frequency and damping rate
\begin{align}
    \Omega_0^2&=[\Bem(\nb+1)+\Babs\nb]\kappa+\Bem \Meb \GDown/\nb,\label{eq:Omega0}\\
    \Gamma &=\Bem(\nb+1)+\Babs\nb+\Bem \Meb/\nb+\GDown\label{eq:Gamma}
\end{align}
depend on the system parameters and the average condensate population $\nb$; for a detailed derivation, see the SI~\cite{supplementalMaterial}. The forcing amplitude is given by $F_0=R_0\Mgb[\Bem(\nb+1)+\Babs\nb]$, where $\Mgb=M-\Meb$ and $M$ is the total molecule number. Solving Eq.~\eqref{eq:drivenBECdiffEq} gives the absolute value of the expected response spectrum
\begin{equation}\label{eq:absBECsol}
   |n_0(\Omega)|=\frac{F_0}{\sqrt{(\Omega^2-\Omega_0^2)^2+\Gamma^2\Omega^2}}.
\end{equation}
Intuitively, this resonance can be understood to result from an effective restoring force that is activated whenever the driven-dissipative system is displaced from its steady state: For $n(t)>\bar n$, cavity loss reduces the photon number with the restoring force proportional to the displacement $n(t)-\bar{n}$; on the other hand, for $n(t)<\bar n$, the pumping acts in the opposite direction and restores the rest 'position' $\bar n$ of the oscillator. For large photon numbers, the zero-frequency response ${|n_0(0)|}/{R_0}={M}/{\kappa}$ depends only on the total molecule number and the cavity losses. The relative phase between BEC and drive is $\phi(\Omega)=-\arccos[(\Omega_0^2-\Omega^2)/\sqrt{(\Omega^2-\Omega_0^2)^2+\Gamma^2\Omega^2}]$. On resonance, $\Omega_\mathrm{r} = \sqrt{\Omega_0^2-\Gamma^2/2}$, one expects the maximum response and a phase shift of $-\pi/2$ because here most of the driving energy is absorbed by the system. The imaginary and real parts of the response are obtained from $\mathrm{Im}\thinspace n_0(\Omega)=-|n_0(\Omega)|\sin \phi$ and $\mathrm{Re}\thinspace n_0(\Omega)=|n_0(\Omega)|\cos\phi$.

Figure~\ref{fig:1}(c) shows three exemplary time traces of the measured response of the photon BEC population (red diamonds) and the drive (blue circles) for frequencies $\Omega/2\pi = \{100,200,300\}\SI{}{\mega\hertz}$. In all cases, the condensate population oscillates at the drive frequency. As the frequency is increased, the phase delay between the leading drive and the trailing BEC response grows from approximately 0 via $-\pi/2$ to $-\pi$ and the response amplitude is largest for $\SI{200}{\mega\hertz}$, which experimentally confirms the resonance of the driven system expected from the model prediction.

\begin{figure}[b]
    \centering
   \includegraphics[width=1.0\columnwidth]{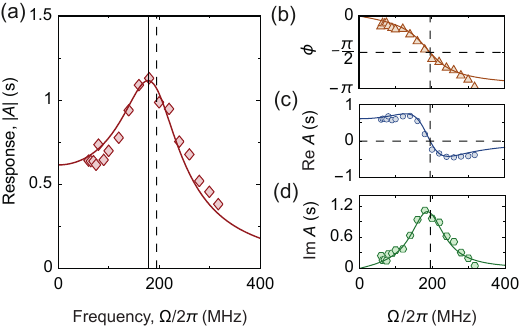}
\caption{Driven optical quantum gas spectroscopy. (a) Response amplitude of the condensate $|A(\Omega)|$ as a function of the driving frequency for a ground state population $\nb=9000$ shows a resonance at $\sqrt{\Omega_0^2-\Gamma^2/2}/2\pi=\SI{178\pm 8}{\mega\hertz}$ (solid vertical line). The dashed line indicates $\Omega_0$. (b) Relative phase $\phi(\Omega)$ between the driving of the bath and the response of the condensate crosses over from $0$ to $-\pi$. At resonance, $\phi=-\pi/2$, within experimental uncertainties. (c) Real and (d) imaginary parts of the response show the resonant behavior of the system as a zero-crossing in $\mathrm{Re}\ A(\Omega)$ and as a peak in $\mathrm{Im}\ A(\Omega)$, respectively. In all panels, solid lines show the theory prediction using the same set of fit parameters for all plots: $\Omega_0/2\pi=\SI{195\pm 7}{\mega\hertz}$, $\Gamma/2\pi=\SI{112\pm 9}{\mega\hertz}$, and $F_0/R_0=\SI{9 \pm 1}{\times 10^{17} \second^{-1}}$. Error bars are smaller than the point size.}
    \label{fig:2}
\end{figure}

\begin{figure*}[t]
    \centering
   \includegraphics[width=1.0\textwidth]{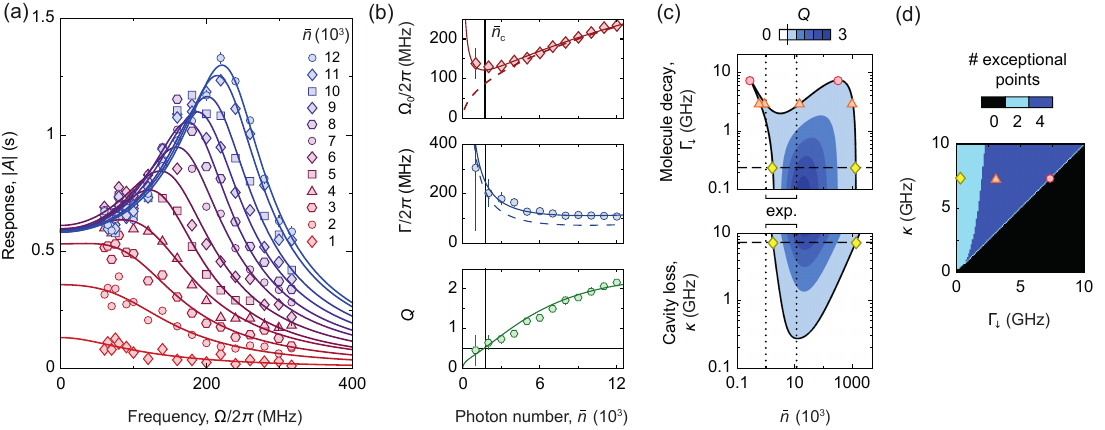}
\caption{Response spectra for increasing condensate population. (a) Amplitude response $|A(\Omega)|$ as a function of the driving frequency for average ground state populations between $\nb=1000$ and $12000$, along with fits (lines). For increasing $\nb$, the spectral profiles exhibit a transition from overdamped to underdamped behavior. (b) Extracted free oscillation frequency $\Omega_0$ (top panel) and damping rate $\Gamma$ (middle) versus $\nb$ with fits (solid lines). For comparison, also theory curves for $\GDown=0$ are shown (dashed lines). The quality factor $Q$ (bottom) shows the transition of the BEC dynamics at $\nb_\mathrm{c} \approx 1790$ photons, where $Q=1/2$ (black lines). Error bars show standard statistical errors. (c) Phase diagram of the driven--dissipative condensate calculated as a function of average photon number $\nb$ and spontaneous loss rate $\GDown$ (top panel, for $\kappa=\SI{7.4}{\giga\hertz}$) or cavity losses $\kappa$ (bottom panel, for $\GDown=\SI{244}{\mega\hertz}$). Colors give the quality factor, where $Q<1/2$ and $Q>1/2$ indicate regions of overdamped and underdamped BEC--bath dynamics, respectively; the dashed lines show the values $\GDown$ and $\kappa$ used in the experiment, and also the experimentally accessed region of average photon numbers is indicated (dotted). In the top panel near $\GDown=\kappa$ and $\nb \approx 300$, a cusp singularity is visible. (d) Number of transitions (exceptional points) in (c) as a function of cavity and molecule loss.}
    \label{fig:3}
\end{figure*}

{\itshape{Driven quantum gas spectroscopy.---}} We perform spectroscopy of the optical quantum gas by driving the bath at different frequencies $\Omega$ for a fixed average photon number $\nb=9000$. Figure~\ref{fig:2}(a) shows the absolute value of the response spectrum normalized by the driving amplitude, $|A(\Omega)|$ with $A(\Omega)=n_0(\Omega)/R_0$, The response amplitude is obtained by fitting the measured time traces of the condensate population, as in the example data from Fig.~\ref{fig:1}(c) and shown in the SI~\cite{supplementalMaterial}. The value of the driving amplitude $R_0$ is calibrated for large photon numbers by using the condition $|n_0(0)|/R_0=M/\kappa$ which follows from Eq.~\eqref{eq:absBECsol}, with here $R_0\approx |n_0(0)|\times \SI{1.7}{\per\second}$. In addition, all data points have been normalized to the respective driving amplitudes at different $\Omega$ to account for residual variations of the AOM diffraction efficiency, also shown in~\cite{supplementalMaterial}. The spectral data (symbols) exhibits a characteristic driven--damped harmonic oscillator spectral shape, and fitting with Eq.~\eqref{eq:absBECsol} yields the free oscillation frequency $\Omega_0/2\pi=\SI{195\pm 7}{\mega\hertz}$, damping rate $\Gamma/2\pi=\SI{112\pm 9}{\mega\hertz}$, and quality factor $Q=\Omega_0/\Gamma=\SI{1.7\pm 0.2}{}$, well consistent with the theory prediction $\Omega_0/2\pi = \SI{204}{\mega\hertz}$ and $\Gamma/2\pi = \SI{110}{\mega\hertz}$ for the experimental values $\GDown=\SI{244}{\mega\hertz}$, $\kappa=\SI{7.4}{\giga\hertz}$, and $\Meb\approx 9\times 10^7$. Figure~\ref{fig:2}(b) gives the extracted phase delay between the drive and the BEC. At small frequencies, the photon number time trace essentially follows the drive phase, while at large frequencies, the phase delay asymptotically reaches $-\pi$ and the condensate lags behind the drive by half a period. From the measured absolute response and the phase, the imaginary and real parts of the response spectrum are extracted, as shown in Figs.~\ref{fig:2}(c) and (d). The imaginary part represents the driving power dissipated by the system, while the real part indicates how closely the condensate follows the drive.

{\itshape{Emergence of driven--dissipative resonance.---}} Next we extend our bath-driving spectroscopy to different condensate populations $\nb$, allowing us to tune the driven-dissipative system between regions with distinct dynamics~\cite{Oeztuerk:2021}. For this, the pump power is varied and the relative modulation depth is kept fixed at roughly $R_0/\GUp= \SI{0.7\pm 0.1}{}\%$ in all experiments, resulting in a relative variation of the condensate population of $\SI{22\pm 11}{}\%$, which is still small enough to remain in the regime of linear response where anharmonic contributions can be neglected~\cite{supplementalMaterial}. Figure~\ref{fig:3}(a) shows the measured response spectra $|A(\Omega)|$ for average photons numbers ranging from $\nb=1000$ to $12000$, along with fits. For small photon numbers $\nb$, we observe overdamped response spectra characterized by a maximum response at $\Omega=0$. At larger photon numbers, a maximum of the response at nonzero frequency $\Omega$ emerges, indicating a transition from the overdamped to an underdamped regime, and a shift of the resonance to larger frequencies is observed. The zero-frequency response for small $\nb$ deviates from $M/\kappa$ due to the molecular losses, while for larger $\nb$ it saturates. The transition point between both regimes, defined by the condition $\Omega_0 = \Gamma/2$, is extracted from the fitted resonance frequencies $\Omega_0(\nb)$ and damping rates $\Gamma(\nb)$. 

Figure~\ref{fig:3}(b) shows the corresponding data along with fits of Eqs.~\eqref{eq:Omega0} and \eqref{eq:Gamma} as a function of the condensate occupancy. For comparison, we also give theory curves for $\GDown=0$, which deviate from the data at small $\nb$. This confirms that excitation losses from spontaneous emission into unconfined modes must be taken into account to adequately describe the system. From the fit we obtain the cavity decay rate $\kappa=\SI{7.4\pm 0.1}{\giga\hertz}$ and the molecule number $M=\SI{4.4\pm 0.3}{}\times 10^9$, consistent with previous results~\cite{Sazhin:2024}. The quality factor shown in the bottom panel of Fig.~\ref{fig:3}(b) shows the transition between overdamped ($Q<1/2$) and underdamped ($Q>1/2$) dynamics.

{\itshape{Open-system phase diagram.---}} The transition between the different BEC--bath dynamics can be further theoretically examined as a function of $\nb$, $\kappa$, and $\GDown$. Figure~\ref{fig:3}(c) shows the phase diagram of the driven condensate versus photon number $\nb$ and spontaneous losses $\GDown$ with overdamped (white region, $Q<1/2$) and underdamped dynamics (blue, $Q>1/2$), where $\kappa, \Babs, \Bem$, and $M$ are kept fixed. Specifically, for $\GDown=\SI{244}{\mega\hertz}$, the condensate is in the underdamped phase between $\nb\approx 1700$ and $\nb\approx 1.2\times 10^6$ photons (yellow diamonds). At the transition the system crosses an exceptional point, where the two eigenvalues of the non-Hermitian matrix describing the photon and excited molecule number dynamics around the steady state $\nb$ and $\bar{\Me}$ coalesce~\cite{Oeztuerk:2021,Sazhin:2024}; thus, the exceptional points considered in the present work are related to the number dynamics and not to the Hamiltonian. Interestingly, at larger spontaneous losses $\GDown$, the BEC dynamics is expected to exhibit four consecutive transitions between the overdamped and underdamped regimes for increased photon number (orange triangles). In particular, at $\nb\approx 300$ and $\GDown=\kappa$, a cusp is visible [Fig.~\ref{fig:3}(c) top, left red circle], which may exhibit a modified scaling behavior with $\nb$~\cite{Rahmani:2024}; this so far unobserved feature should be investigated further in the future by using cavity media with larger losses. The bottom panel of Fig.~\ref{fig:3}(c) depicts the corresponding phase diagram for fixed $\GDown$ that exhibits the underdamped regime for sufficiently large cavity losses $\kappa$. This can be understood from a stronger effective restoring force back to the steady state $\nb$ after a photon number displacement, resulting in a larger oscillation frequency. Figure~\ref{fig:3}(d) summarizes the number of predicted exceptional points for a set of cavity loss and molecule decay rates. At the experimental parameters (yellow diamond), two exceptional points are expected~\cite{Sazhin:2024,Bode:2024}; note that the one at $\nb\approx 1790$ has been observed experimentally. At larger $\GDown$ four points emerge. For the cusp condition ($\GDown=\kappa$, red circle) two exceptional points are crossed, and for $\GDown>\kappa$ the system remains overdamped.

{\itshape{Conclusion.---}} We have determined the spectral particle number response of a photon BEC inside a dye-filled microcavity under periodic driving of its molecule bath, in this way resolving the excitation spectrum of the coupled photon-molecule system. The corresponding elementary excitation is understood as a collective state of photons and molecules in the weak coupling regime, where the molecules drag the photons behind them as they seek to adjust to the modified chemical potential. Throughout the driving, the condensate remains in quasi-equilibrium because thermalization occurs much faster than losses. The spectral shape of the response for weak driving much resembles that of a driven--damped harmonic oscillator, exhibiting a transition from an overdamped to an underdamped regime as the mean condensate occupancy is increased. Dissipation from spontaneous molecule decay and cavity losses here play an important role in the BEC--bath dynamics and both effects act in opposite directions: Large spontaneous losses induce overdamped dynamics, while large cavity losses suppress them. From a more general viewpoint, the agreement between the theoretical predictions and the experimental results establishes single-frequency reservoir driving as a spectroscopic method to access the dynamic properties of optical quantum gases.

Our method paves the way for testing the frequency-dependent fluctuation--dissipation theorem of the optical condensate~\cite{Chiocchetta:2017b,Oeztuerk:2023}. A future experimental challenge to explore the predicted cusp in the non-Hermitian phase diagram is the realization of tunable photon and molecule losses using different cavities and emitter materials. For this, it will be interesting to test whether the recently observed semiconductor-based photon BECs~\cite{Schofield:2024,Pieczarka:2024} exhibit a similar excitation spectrum as the one observed here. Further prospects include the study of sound propagation in optical quantum gases under periodic driving in spatially extended geometries, such as box or lattice potentials~\cite{Estrecho:2021,Busley:2022}. Exploring the interplay between coherent tunneling and incoherent bath-mediated transport in driven--dissipative arrays of coupled cavities using the demonstrated bath driving spectroscopy offers new possibilities to study unexplored non-Hermitian topological effects~\cite{Wanjura:2020}, such as the bosonic skin effect~\cite{Garbe:2024}.

\vspace{0.5cm}

{\itshape{Acknowledgements.---}} We thank A. Buchleitner, A. Rahmani, A. Redmann, and K. Karkihalli Umesh for fruitful discussions. This work was financially supported by the DFG within SFB/TR 185 (277625399) and the Cluster of Excellence ML4Q (EXC 2004/1–390534769). J.S. acknowledges support by the EU (ERC, TopoGrand, 101040409). A.E. has received funding from the European Union's Horizon 2020 research and innovation program under the Marie Sk\l odowska-Curie grant agreement (847471).

\vspace{0.5cm}

{\itshape{Data availability.---}}  The data that support the findings of this Letter are openly available~\cite{Erglis:2025data}.


\begin{thebibliography}{45}%
\makeatletter
\providecommand \@ifxundefined [1]{%
 \@ifx{#1\undefined}
}%
\providecommand \@ifnum [1]{%
 \ifnum #1\expandafter \@firstoftwo
 \else \expandafter \@secondoftwo
 \fi
}%
\providecommand \@ifx [1]{%
 \ifx #1\expandafter \@firstoftwo
 \else \expandafter \@secondoftwo
 \fi
}%
\providecommand \natexlab [1]{#1}%
\providecommand \enquote  [1]{``#1''}%
\providecommand \bibnamefont  [1]{#1}%
\providecommand \bibfnamefont [1]{#1}%
\providecommand \citenamefont [1]{#1}%
\providecommand \href@noop [0]{\@secondoftwo}%
\providecommand \href [0]{\begingroup \@sanitize@url \@href}%
\providecommand \@href[1]{\@@startlink{#1}\@@href}%
\providecommand \@@href[1]{\endgroup#1\@@endlink}%
\providecommand \@sanitize@url [0]{\catcode `\\12\catcode `\$12\catcode
  `\&12\catcode `\#12\catcode `\^12\catcode `\_12\catcode `\%12\relax}%
\providecommand \@@startlink[1]{}%
\providecommand \@@endlink[0]{}%
\providecommand \url  [0]{\begingroup\@sanitize@url \@url }%
\providecommand \@url [1]{\endgroup\@href {#1}{\urlprefix }}%
\providecommand \urlprefix  [0]{URL }%
\providecommand \Eprint [0]{\href }%
\providecommand \doibase [0]{https://doi.org/}%
\providecommand \selectlanguage [0]{\@gobble}%
\providecommand \bibinfo  [0]{\@secondoftwo}%
\providecommand \bibfield  [0]{\@secondoftwo}%
\providecommand \translation [1]{[#1]}%
\providecommand \BibitemOpen [0]{}%
\providecommand \bibitemStop [0]{}%
\providecommand \bibitemNoStop [0]{.\EOS\space}%
\providecommand \EOS [0]{\spacefactor3000\relax}%
\providecommand \BibitemShut  [1]{\csname bibitem#1\endcsname}%
\let\auto@bib@innerbib\@empty
\bibitem [{\citenamefont {Oka}\ and\ \citenamefont
  {Kitamura}(2019)}]{Oka:2019}%
  \BibitemOpen
  \bibfield  {author} {\bibinfo {author} {\bibfnamefont {T.}~\bibnamefont
  {Oka}}\ and\ \bibinfo {author} {\bibfnamefont {S.}~\bibnamefont {Kitamura}},\
  }\bibfield  {title} {\bibinfo {title} {Floquet engineering of quantum
  materials},\ }\href
  {https://doi.org/10.1146/annurev-conmatphys-031218-013423} {\bibfield
  {journal} {\bibinfo  {journal} {Annu. Rev. Condens. Matter Phys.}\ }\textbf
  {\bibinfo {volume} {10}},\ \bibinfo {pages} {387} (\bibinfo {year}
  {2019})}\BibitemShut {NoStop}%
\bibitem [{\citenamefont {Eckardt}(2017)}]{Eckardt:2017}%
  \BibitemOpen
  \bibfield  {author} {\bibinfo {author} {\bibfnamefont {A.}~\bibnamefont
  {Eckardt}},\ }\bibfield  {title} {\bibinfo {title} {Colloquium: Atomic
  quantum gases in periodically driven optical lattices},\ }\href
  {https://doi.org/10.1103/RevModPhys.89.011004} {\bibfield  {journal}
  {\bibinfo  {journal} {Rev. Mod. Phys.}\ }\textbf {\bibinfo {volume} {89}},\
  \bibinfo {pages} {011004} (\bibinfo {year} {2017})}\BibitemShut {NoStop}%
\bibitem [{\citenamefont {Holthaus}(2015)}]{Holthaus:2015}%
  \BibitemOpen
  \bibfield  {author} {\bibinfo {author} {\bibfnamefont {M.}~\bibnamefont
  {Holthaus}},\ }\bibfield  {title} {\bibinfo {title} {Floquet engineering with
  quasienergy bands of periodically driven optical lattices},\ }\href
  {https://doi.org/10.1088/0953-4075/49/1/013001} {\bibfield  {journal}
  {\bibinfo  {journal} {J. Phys. B: At., Mol. Opt. Phys.}\ }\textbf {\bibinfo
  {volume} {49}},\ \bibinfo {pages} {013001} (\bibinfo {year}
  {2015})}\BibitemShut {NoStop}%
\bibitem [{\citenamefont {Castro}\ \emph {et~al.}(2022)\citenamefont {Castro},
  \citenamefont {De~Giovannini}, \citenamefont {Sato}, \citenamefont
  {H\"ubener},\ and\ \citenamefont {Rubio}}]{Castro:2022}%
  \BibitemOpen
  \bibfield  {author} {\bibinfo {author} {\bibfnamefont {A.}~\bibnamefont
  {Castro}}, \bibinfo {author} {\bibfnamefont {U.}~\bibnamefont
  {De~Giovannini}}, \bibinfo {author} {\bibfnamefont {S.~A.}\ \bibnamefont
  {Sato}}, \bibinfo {author} {\bibfnamefont {H.}~\bibnamefont {H\"ubener}},\
  and\ \bibinfo {author} {\bibfnamefont {A.}~\bibnamefont {Rubio}},\ }\bibfield
   {title} {\bibinfo {title} {Floquet engineering the band structure of
  materials with optimal control theory},\ }\href
  {https://doi.org/10.1103/PhysRevResearch.4.033213} {\bibfield  {journal}
  {\bibinfo  {journal} {Phys. Rev. Res.}\ }\textbf {\bibinfo {volume} {4}},\
  \bibinfo {pages} {033213} (\bibinfo {year} {2022})}\BibitemShut {NoStop}%
\bibitem [{\citenamefont {Rechtsman}\ \emph {et~al.}(2013)\citenamefont
  {Rechtsman}, \citenamefont {Zeuner}, \citenamefont {Plotnik}, \citenamefont
  {Lumer}, \citenamefont {Podolsky}, \citenamefont {Dreisow}, \citenamefont
  {Nolte}, \citenamefont {Segev},\ and\ \citenamefont
  {Szameit}}]{Rechtsman:2013}%
  \BibitemOpen
  \bibfield  {author} {\bibinfo {author} {\bibfnamefont {M.~C.}\ \bibnamefont
  {Rechtsman}}, \bibinfo {author} {\bibfnamefont {J.~M.}\ \bibnamefont
  {Zeuner}}, \bibinfo {author} {\bibfnamefont {Y.}~\bibnamefont {Plotnik}},
  \bibinfo {author} {\bibfnamefont {Y.}~\bibnamefont {Lumer}}, \bibinfo
  {author} {\bibfnamefont {D.}~\bibnamefont {Podolsky}}, \bibinfo {author}
  {\bibfnamefont {F.}~\bibnamefont {Dreisow}}, \bibinfo {author} {\bibfnamefont
  {S.}~\bibnamefont {Nolte}}, \bibinfo {author} {\bibfnamefont
  {M.}~\bibnamefont {Segev}},\ and\ \bibinfo {author} {\bibfnamefont
  {A.}~\bibnamefont {Szameit}},\ }\bibfield  {title} {\bibinfo {title}
  {Photonic floquet topological insulators},\ }\href
  {https://doi.org/10.1038/nature12066} {\bibfield  {journal} {\bibinfo
  {journal} {Nature}\ }\textbf {\bibinfo {volume} {496}},\ \bibinfo {pages}
  {196} (\bibinfo {year} {2013})}\BibitemShut {NoStop}%
\bibitem [{\citenamefont {Jotzu}\ \emph {et~al.}(2014)\citenamefont {Jotzu},
  \citenamefont {Messer}, \citenamefont {Desbuquois}, \citenamefont {Lebrat},
  \citenamefont {Uehlinger}, \citenamefont {Greif},\ and\ \citenamefont
  {Esslinger}}]{Jotzu:2014}%
  \BibitemOpen
  \bibfield  {author} {\bibinfo {author} {\bibfnamefont {G.}~\bibnamefont
  {Jotzu}}, \bibinfo {author} {\bibfnamefont {M.}~\bibnamefont {Messer}},
  \bibinfo {author} {\bibfnamefont {R.}~\bibnamefont {Desbuquois}}, \bibinfo
  {author} {\bibfnamefont {M.}~\bibnamefont {Lebrat}}, \bibinfo {author}
  {\bibfnamefont {T.}~\bibnamefont {Uehlinger}}, \bibinfo {author}
  {\bibfnamefont {D.}~\bibnamefont {Greif}},\ and\ \bibinfo {author}
  {\bibfnamefont {T.}~\bibnamefont {Esslinger}},\ }\bibfield  {title} {\bibinfo
  {title} {Experimental realization of the topological {H}aldane model with
  ultracold fermions},\ }\href {https://doi.org/10.1038/nature13915} {\bibfield
   {journal} {\bibinfo  {journal} {Nature}\ }\textbf {\bibinfo {volume}
  {515}},\ \bibinfo {pages} {237} (\bibinfo {year} {2014})}\BibitemShut
  {NoStop}%
\bibitem [{\citenamefont {Clark}\ \emph {et~al.}(2019)\citenamefont {Clark},
  \citenamefont {Jia}, \citenamefont {Schine}, \citenamefont {Baum},
  \citenamefont {Georgakopoulos},\ and\ \citenamefont {Simon}}]{Clark:2019}%
  \BibitemOpen
  \bibfield  {author} {\bibinfo {author} {\bibfnamefont {L.~W.}\ \bibnamefont
  {Clark}}, \bibinfo {author} {\bibfnamefont {N.}~\bibnamefont {Jia}}, \bibinfo
  {author} {\bibfnamefont {N.}~\bibnamefont {Schine}}, \bibinfo {author}
  {\bibfnamefont {C.}~\bibnamefont {Baum}}, \bibinfo {author} {\bibfnamefont
  {A.}~\bibnamefont {Georgakopoulos}},\ and\ \bibinfo {author} {\bibfnamefont
  {J.}~\bibnamefont {Simon}},\ }\bibfield  {title} {\bibinfo {title}
  {Interacting {F}loquet polaritons},\ }\href
  {https://doi.org/10.1038/s41586-019-1354-5} {\bibfield  {journal} {\bibinfo
  {journal} {Nature}\ }\textbf {\bibinfo {volume} {571}},\ \bibinfo {pages}
  {532} (\bibinfo {year} {2019})}\BibitemShut {NoStop}%
\bibitem [{\citenamefont {Ikeda}\ and\ \citenamefont
  {Sato}(2020)}]{Ikeda:2020}%
  \BibitemOpen
  \bibfield  {author} {\bibinfo {author} {\bibfnamefont {T.~N.}\ \bibnamefont
  {Ikeda}}\ and\ \bibinfo {author} {\bibfnamefont {M.}~\bibnamefont {Sato}},\
  }\bibfield  {title} {\bibinfo {title} {General description for nonequilibrium
  steady states in periodically driven dissipative quantum systems},\ }\href
  {https://doi.org/10.1126/sciadv.abb4019} {\bibfield  {journal} {\bibinfo
  {journal} {Sci. Adv.}\ }\textbf {\bibinfo {volume} {6}},\ \bibinfo {pages}
  {eabb4019} (\bibinfo {year} {2020})}\BibitemShut {NoStop}%
\bibitem [{\citenamefont {Sato}\ \emph {et~al.}(2020)\citenamefont {Sato},
  \citenamefont {Giovannini}, \citenamefont {Aeschlimann}, \citenamefont
  {Gierz}, \citenamefont {Hübener},\ and\ \citenamefont {Rubio}}]{Sato:2020}%
  \BibitemOpen
  \bibfield  {author} {\bibinfo {author} {\bibfnamefont {S.~A.}\ \bibnamefont
  {Sato}}, \bibinfo {author} {\bibfnamefont {U.~D.}\ \bibnamefont
  {Giovannini}}, \bibinfo {author} {\bibfnamefont {S.}~\bibnamefont
  {Aeschlimann}}, \bibinfo {author} {\bibfnamefont {I.}~\bibnamefont {Gierz}},
  \bibinfo {author} {\bibfnamefont {H.}~\bibnamefont {Hübener}},\ and\
  \bibinfo {author} {\bibfnamefont {A.}~\bibnamefont {Rubio}},\ }\bibfield
  {title} {\bibinfo {title} {Floquet states in dissipative open quantum
  systems},\ }\href {https://doi.org/10.1088/1361-6455/abb127} {\bibfield
  {journal} {\bibinfo  {journal} {J. Phys. B: At. Mol. Opt. Phys.}\ }\textbf
  {\bibinfo {volume} {53}},\ \bibinfo {pages} {225601} (\bibinfo {year}
  {2020})}\BibitemShut {NoStop}%
\bibitem [{\citenamefont {Ochoa}(2024)}]{Ochoa:2024}%
  \BibitemOpen
  \bibfield  {author} {\bibinfo {author} {\bibfnamefont {M.~A.}\ \bibnamefont
  {Ochoa}},\ }\href {https://arxiv.org/abs/2402.18560} {\bibinfo {title} {Lossy
  anharmonic polaritons under periodic driving}} (\bibinfo {year} {2024}),\
  \Eprint {https://arxiv.org/abs/2402.18560} {arXiv:2402.18560 [quant-ph]}
  \BibitemShut {NoStop}%
\bibitem [{\citenamefont {Barreiro}\ \emph {et~al.}(2011)\citenamefont
  {Barreiro}, \citenamefont {M{\"u}ller}, \citenamefont {Schindler},
  \citenamefont {Nigg}, \citenamefont {Monz}, \citenamefont {Chwalla},
  \citenamefont {Hennrich}, \citenamefont {Roos}, \citenamefont {Zoller},\ and\
  \citenamefont {Blatt}}]{Barreiro:2011}%
  \BibitemOpen
  \bibfield  {author} {\bibinfo {author} {\bibfnamefont {J.~T.}\ \bibnamefont
  {Barreiro}}, \bibinfo {author} {\bibfnamefont {M.}~\bibnamefont
  {M{\"u}ller}}, \bibinfo {author} {\bibfnamefont {P.}~\bibnamefont
  {Schindler}}, \bibinfo {author} {\bibfnamefont {D.}~\bibnamefont {Nigg}},
  \bibinfo {author} {\bibfnamefont {T.}~\bibnamefont {Monz}}, \bibinfo {author}
  {\bibfnamefont {M.}~\bibnamefont {Chwalla}}, \bibinfo {author} {\bibfnamefont
  {M.}~\bibnamefont {Hennrich}}, \bibinfo {author} {\bibfnamefont {C.~F.}\
  \bibnamefont {Roos}}, \bibinfo {author} {\bibfnamefont {P.}~\bibnamefont
  {Zoller}},\ and\ \bibinfo {author} {\bibfnamefont {R.}~\bibnamefont
  {Blatt}},\ }\bibfield  {title} {\bibinfo {title} {An open-system quantum
  simulator with trapped ions},\ }\href {https://doi.org/10.1038/nature09801}
  {\bibfield  {journal} {\bibinfo  {journal} {Nature}\ }\textbf {\bibinfo
  {volume} {470}},\ \bibinfo {pages} {486} (\bibinfo {year}
  {2011})}\BibitemShut {NoStop}%
\bibitem [{\citenamefont {Barontini}\ \emph {et~al.}(2013)\citenamefont
  {Barontini}, \citenamefont {Labouvie}, \citenamefont {Stubenrauch},
  \citenamefont {Vogler}, \citenamefont {Guarrera},\ and\ \citenamefont
  {Ott}}]{Barontini:2013}%
  \BibitemOpen
  \bibfield  {author} {\bibinfo {author} {\bibfnamefont {G.}~\bibnamefont
  {Barontini}}, \bibinfo {author} {\bibfnamefont {R.}~\bibnamefont {Labouvie}},
  \bibinfo {author} {\bibfnamefont {F.}~\bibnamefont {Stubenrauch}}, \bibinfo
  {author} {\bibfnamefont {A.}~\bibnamefont {Vogler}}, \bibinfo {author}
  {\bibfnamefont {V.}~\bibnamefont {Guarrera}},\ and\ \bibinfo {author}
  {\bibfnamefont {H.}~\bibnamefont {Ott}},\ }\bibfield  {title} {\bibinfo
  {title} {Controlling the dynamics of an open many-body quantum system with
  localized dissipation},\ }\href
  {https://doi.org/10.1103/PhysRevLett.110.035302} {\bibfield  {journal}
  {\bibinfo  {journal} {Phys. Rev. Lett.}\ }\textbf {\bibinfo {volume} {110}},\
  \bibinfo {pages} {035302} (\bibinfo {year} {2013})}\BibitemShut {NoStop}%
\bibitem [{\citenamefont {Tomita}\ \emph {et~al.}(2017)\citenamefont {Tomita},
  \citenamefont {Nakajima}, \citenamefont {Danshita}, \citenamefont {Takasu},\
  and\ \citenamefont {Takahashi}}]{Tomita:2017}%
  \BibitemOpen
  \bibfield  {author} {\bibinfo {author} {\bibfnamefont {T.}~\bibnamefont
  {Tomita}}, \bibinfo {author} {\bibfnamefont {S.}~\bibnamefont {Nakajima}},
  \bibinfo {author} {\bibfnamefont {I.}~\bibnamefont {Danshita}}, \bibinfo
  {author} {\bibfnamefont {Y.}~\bibnamefont {Takasu}},\ and\ \bibinfo {author}
  {\bibfnamefont {Y.}~\bibnamefont {Takahashi}},\ }\bibfield  {title} {\bibinfo
  {title} {Observation of the {M}ott insulator to superfluid crossover of a
  driven-dissipative {B}ose-{H}ubbard system},\ }\href
  {https://doi.org/10.1126/sciadv.1701513} {\bibfield  {journal} {\bibinfo
  {journal} {Sci. Adv.}\ }\textbf {\bibinfo {volume} {3}},\ \bibinfo {pages}
  {e1701513} (\bibinfo {year} {2017})}\BibitemShut {NoStop}%
\bibitem [{\citenamefont {Wetter}\ \emph {et~al.}(2023)\citenamefont {Wetter},
  \citenamefont {Fleischhauer}, \citenamefont {Linden},\ and\ \citenamefont
  {Schmitt}}]{Wetter:2023}%
  \BibitemOpen
  \bibfield  {author} {\bibinfo {author} {\bibfnamefont {H.}~\bibnamefont
  {Wetter}}, \bibinfo {author} {\bibfnamefont {M.}~\bibnamefont
  {Fleischhauer}}, \bibinfo {author} {\bibfnamefont {S.}~\bibnamefont
  {Linden}},\ and\ \bibinfo {author} {\bibfnamefont {J.}~\bibnamefont
  {Schmitt}},\ }\bibfield  {title} {\bibinfo {title} {Observation of a
  topological edge state stabilized by dissipation},\ }\href
  {https://doi.org/10.1103/PhysRevLett.131.083801} {\bibfield  {journal}
  {\bibinfo  {journal} {Phys. Rev. Lett.}\ }\textbf {\bibinfo {volume} {131}},\
  \bibinfo {pages} {083801} (\bibinfo {year} {2023})}\BibitemShut {NoStop}%
\bibitem [{\citenamefont {Fedorova}\ \emph {et~al.}(2020)\citenamefont
  {Fedorova}, \citenamefont {Qiu}, \citenamefont {Linden},\ and\ \citenamefont
  {Kroha}}]{Fedorova:2020}%
  \BibitemOpen
  \bibfield  {author} {\bibinfo {author} {\bibfnamefont {Z.}~\bibnamefont
  {Fedorova}}, \bibinfo {author} {\bibfnamefont {H.}~\bibnamefont {Qiu}},
  \bibinfo {author} {\bibfnamefont {S.}~\bibnamefont {Linden}},\ and\ \bibinfo
  {author} {\bibfnamefont {J.}~\bibnamefont {Kroha}},\ }\bibfield  {title}
  {\bibinfo {title} {Observation of topological transport quantization by
  dissipation in fast {T}houless pumps},\ }\href
  {https://doi.org/10.1038/s41467-020-17510-z} {\bibfield  {journal} {\bibinfo
  {journal} {Nat. Commun.}\ }\textbf {\bibinfo {volume} {11}},\ \bibinfo
  {pages} {3758} (\bibinfo {year} {2020})}\BibitemShut {NoStop}%
\bibitem [{\citenamefont {Gnusov}\ \emph {et~al.}(2023)\citenamefont {Gnusov},
  \citenamefont {Harrison}, \citenamefont {Alyatkin}, \citenamefont {Sitnik},
  \citenamefont {Töpfer}, \citenamefont {Sigurdsson},\ and\ \citenamefont
  {Lagoudakis}}]{Gnusov:2023}%
  \BibitemOpen
  \bibfield  {author} {\bibinfo {author} {\bibfnamefont {I.}~\bibnamefont
  {Gnusov}}, \bibinfo {author} {\bibfnamefont {S.}~\bibnamefont {Harrison}},
  \bibinfo {author} {\bibfnamefont {S.}~\bibnamefont {Alyatkin}}, \bibinfo
  {author} {\bibfnamefont {K.}~\bibnamefont {Sitnik}}, \bibinfo {author}
  {\bibfnamefont {J.}~\bibnamefont {Töpfer}}, \bibinfo {author} {\bibfnamefont
  {H.}~\bibnamefont {Sigurdsson}},\ and\ \bibinfo {author} {\bibfnamefont
  {P.}~\bibnamefont {Lagoudakis}},\ }\bibfield  {title} {\bibinfo {title}
  {Quantum vortex formation in the “rotating bucket” experiment with
  polariton condensates},\ }\href {https://doi.org/10.1126/sciadv.add1299}
  {\bibfield  {journal} {\bibinfo  {journal} {Sci. Adv.}\ }\textbf {\bibinfo
  {volume} {9}},\ \bibinfo {pages} {eadd1299} (\bibinfo {year}
  {2023})}\BibitemShut {NoStop}%
\bibitem [{\citenamefont {del Valle-Inclan~Redondo}\ \emph
  {et~al.}(2023)\citenamefont {del Valle-Inclan~Redondo}, \citenamefont
  {Schneider}, \citenamefont {Klembt}, \citenamefont {H{\"o}fling},
  \citenamefont {Tarucha},\ and\ \citenamefont {Fraser}}]{delValle:2023}%
  \BibitemOpen
  \bibfield  {author} {\bibinfo {author} {\bibfnamefont {Y.}~\bibnamefont {del
  Valle-Inclan~Redondo}}, \bibinfo {author} {\bibfnamefont {C.}~\bibnamefont
  {Schneider}}, \bibinfo {author} {\bibfnamefont {S.}~\bibnamefont {Klembt}},
  \bibinfo {author} {\bibfnamefont {S.}~\bibnamefont {H{\"o}fling}}, \bibinfo
  {author} {\bibfnamefont {S.}~\bibnamefont {Tarucha}},\ and\ \bibinfo {author}
  {\bibfnamefont {M.~D.}\ \bibnamefont {Fraser}},\ }\bibfield  {title}
  {\bibinfo {title} {Optically driven rotation of exciton--polariton
  condensates},\ }\href {https://doi.org/10.1021/acs.nanolett.3c01021}
  {\bibfield  {journal} {\bibinfo  {journal} {Nano Lett.}\ }\textbf {\bibinfo
  {volume} {23}},\ \bibinfo {pages} {4564} (\bibinfo {year}
  {2023})}\BibitemShut {NoStop}%
\bibitem [{\citenamefont {del Valle Inclan~Redondo}\ \emph
  {et~al.}(2024)\citenamefont {del Valle Inclan~Redondo}, \citenamefont {Xu},
  \citenamefont {Liew}, \citenamefont {Ostrovskaya}, \citenamefont {Stegmaier},
  \citenamefont {Thomale}, \citenamefont {Schneider}, \citenamefont {Dam},
  \citenamefont {Klembt}, \citenamefont {H{\"o}fling}, \citenamefont
  {Tarucha},\ and\ \citenamefont {Fraser}}]{delValle:2024}%
  \BibitemOpen
  \bibfield  {author} {\bibinfo {author} {\bibfnamefont {Y.}~\bibnamefont {del
  Valle Inclan~Redondo}}, \bibinfo {author} {\bibfnamefont {X.}~\bibnamefont
  {Xu}}, \bibinfo {author} {\bibfnamefont {T.~C.~H.}\ \bibnamefont {Liew}},
  \bibinfo {author} {\bibfnamefont {E.~A.}\ \bibnamefont {Ostrovskaya}},
  \bibinfo {author} {\bibfnamefont {A.}~\bibnamefont {Stegmaier}}, \bibinfo
  {author} {\bibfnamefont {R.}~\bibnamefont {Thomale}}, \bibinfo {author}
  {\bibfnamefont {C.}~\bibnamefont {Schneider}}, \bibinfo {author}
  {\bibfnamefont {S.}~\bibnamefont {Dam}}, \bibinfo {author} {\bibfnamefont
  {S.}~\bibnamefont {Klembt}}, \bibinfo {author} {\bibfnamefont
  {S.}~\bibnamefont {H{\"o}fling}}, \bibinfo {author} {\bibfnamefont
  {S.}~\bibnamefont {Tarucha}},\ and\ \bibinfo {author} {\bibfnamefont {M.~D.}\
  \bibnamefont {Fraser}},\ }\bibfield  {title} {\bibinfo {title}
  {Non-reciprocal band structures in an exciton--polariton floquet optical
  lattice},\ }\href {https://doi.org/10.1038/s41566-024-01424-z} {\bibfield
  {journal} {\bibinfo  {journal} {Nat. Photonics}\ }\textbf {\bibinfo {volume}
  {18}},\ \bibinfo {pages} {548} (\bibinfo {year} {2024})}\BibitemShut
  {NoStop}%
\bibitem [{\citenamefont {Klaers}\ \emph {et~al.}(2010)\citenamefont {Klaers},
  \citenamefont {Schmitt}, \citenamefont {Vewinger},\ and\ \citenamefont
  {Weitz}}]{Klaers:2010}%
  \BibitemOpen
  \bibfield  {author} {\bibinfo {author} {\bibfnamefont {J.}~\bibnamefont
  {Klaers}}, \bibinfo {author} {\bibfnamefont {J.}~\bibnamefont {Schmitt}},
  \bibinfo {author} {\bibfnamefont {F.}~\bibnamefont {Vewinger}},\ and\
  \bibinfo {author} {\bibfnamefont {M.}~\bibnamefont {Weitz}},\ }\bibfield
  {title} {\bibinfo {title} {{B}ose--{E}instein condensation of photons in an
  optical microcavity},\ }\href {https://doi.org/10.1038/nature09567}
  {\bibfield  {journal} {\bibinfo  {journal} {Nature}\ }\textbf {\bibinfo
  {volume} {468}},\ \bibinfo {pages} {545} (\bibinfo {year}
  {2010})}\BibitemShut {NoStop}%
\bibitem [{\citenamefont {Marelic}\ and\ \citenamefont
  {Nyman}(2015)}]{Marelic:2015}%
  \BibitemOpen
  \bibfield  {author} {\bibinfo {author} {\bibfnamefont {J.}~\bibnamefont
  {Marelic}}\ and\ \bibinfo {author} {\bibfnamefont {R.~A.}\ \bibnamefont
  {Nyman}},\ }\bibfield  {title} {\bibinfo {title} {Experimental evidence for
  inhomogeneous pumping and energy-dependent effects in photon
  {B}ose-{E}instein condensation},\ }\href
  {https://doi.org/10.1103/PhysRevA.91.033813} {\bibfield  {journal} {\bibinfo
  {journal} {Phys. Rev. A}\ }\textbf {\bibinfo {volume} {91}},\ \bibinfo
  {pages} {033813} (\bibinfo {year} {2015})}\BibitemShut {NoStop}%
\bibitem [{\citenamefont {Greveling}\ \emph {et~al.}(2018)\citenamefont
  {Greveling}, \citenamefont {Perrier},\ and\ \citenamefont {van
  Oosten}}]{Greveling:2018}%
  \BibitemOpen
  \bibfield  {author} {\bibinfo {author} {\bibfnamefont {S.}~\bibnamefont
  {Greveling}}, \bibinfo {author} {\bibfnamefont {K.~L.}\ \bibnamefont
  {Perrier}},\ and\ \bibinfo {author} {\bibfnamefont {D.}~\bibnamefont {van
  Oosten}},\ }\bibfield  {title} {\bibinfo {title} {Density distribution of a
  {B}ose-{E}instein condensate of photons in a dye-filled microcavity},\ }\href
  {https://doi.org/10.1103/PhysRevA.98.013810} {\bibfield  {journal} {\bibinfo
  {journal} {Phys. Rev. A}\ }\textbf {\bibinfo {volume} {98}},\ \bibinfo
  {pages} {013810} (\bibinfo {year} {2018})}\BibitemShut {NoStop}%
\bibitem [{\citenamefont {Schofield}\ \emph {et~al.}(2024)\citenamefont
  {Schofield}, \citenamefont {Fu}, \citenamefont {Clarke}, \citenamefont
  {Farrer}, \citenamefont {Trapalis}, \citenamefont {Dhar}, \citenamefont
  {Mukherjee}, \citenamefont {Severs~Millard}, \citenamefont {Heffernan},
  \citenamefont {Mintert}, \citenamefont {Nyman},\ and\ \citenamefont
  {Oulton}}]{Schofield:2024}%
  \BibitemOpen
  \bibfield  {author} {\bibinfo {author} {\bibfnamefont {R.~C.}\ \bibnamefont
  {Schofield}}, \bibinfo {author} {\bibfnamefont {M.}~\bibnamefont {Fu}},
  \bibinfo {author} {\bibfnamefont {E.}~\bibnamefont {Clarke}}, \bibinfo
  {author} {\bibfnamefont {I.}~\bibnamefont {Farrer}}, \bibinfo {author}
  {\bibfnamefont {A.}~\bibnamefont {Trapalis}}, \bibinfo {author}
  {\bibfnamefont {H.~S.}\ \bibnamefont {Dhar}}, \bibinfo {author}
  {\bibfnamefont {R.}~\bibnamefont {Mukherjee}}, \bibinfo {author}
  {\bibfnamefont {T.}~\bibnamefont {Severs~Millard}}, \bibinfo {author}
  {\bibfnamefont {J.}~\bibnamefont {Heffernan}}, \bibinfo {author}
  {\bibfnamefont {F.}~\bibnamefont {Mintert}}, \bibinfo {author} {\bibfnamefont
  {R.~A.}\ \bibnamefont {Nyman}},\ and\ \bibinfo {author} {\bibfnamefont
  {R.~F.}\ \bibnamefont {Oulton}},\ }\bibfield  {title} {\bibinfo {title}
  {Bose--{E}instein condensation of light in a semiconductor quantum well
  microcavity},\ }\bibfield  {journal} {\bibinfo  {journal} {Nat. Photon.}\
  }\href {https://doi.org/10.1038/s41566-024-01491-2}
  {10.1038/s41566-024-01491-2} (\bibinfo {year} {2024})\BibitemShut {NoStop}%
\bibitem [{\citenamefont {Pieczarka}\ \emph {et~al.}(2024)\citenamefont
  {Pieczarka}, \citenamefont {G{\k{e}}bski}, \citenamefont {Piasecka},
  \citenamefont {Lott}, \citenamefont {Pelster}, \citenamefont {Wasiak},\ and\
  \citenamefont {Czyszanowski}}]{Pieczarka:2024}%
  \BibitemOpen
  \bibfield  {author} {\bibinfo {author} {\bibfnamefont {M.}~\bibnamefont
  {Pieczarka}}, \bibinfo {author} {\bibfnamefont {M.}~\bibnamefont
  {G{\k{e}}bski}}, \bibinfo {author} {\bibfnamefont {A.~N.}\ \bibnamefont
  {Piasecka}}, \bibinfo {author} {\bibfnamefont {J.~A.}\ \bibnamefont {Lott}},
  \bibinfo {author} {\bibfnamefont {A.}~\bibnamefont {Pelster}}, \bibinfo
  {author} {\bibfnamefont {M.}~\bibnamefont {Wasiak}},\ and\ \bibinfo {author}
  {\bibfnamefont {T.}~\bibnamefont {Czyszanowski}},\ }\bibfield  {title}
  {\bibinfo {title} {{B}ose--{E}instein condensation of photons in a
  vertical-cavity surface-emitting laser},\ }\bibfield  {journal} {\bibinfo
  {journal} {Nat. Photon.}\ }\href {https://doi.org/10.1038/s41566-024-01478-z}
  {10.1038/s41566-024-01478-z} (\bibinfo {year} {2024})\BibitemShut {NoStop}%
\bibitem [{\citenamefont {Schmitt}\ \emph {et~al.}(2015)\citenamefont
  {Schmitt}, \citenamefont {Damm}, \citenamefont {Dung}, \citenamefont
  {Vewinger}, \citenamefont {Klaers},\ and\ \citenamefont
  {Weitz}}]{Schmitt:2015}%
  \BibitemOpen
  \bibfield  {author} {\bibinfo {author} {\bibfnamefont {J.}~\bibnamefont
  {Schmitt}}, \bibinfo {author} {\bibfnamefont {T.}~\bibnamefont {Damm}},
  \bibinfo {author} {\bibfnamefont {D.}~\bibnamefont {Dung}}, \bibinfo {author}
  {\bibfnamefont {F.}~\bibnamefont {Vewinger}}, \bibinfo {author}
  {\bibfnamefont {J.}~\bibnamefont {Klaers}},\ and\ \bibinfo {author}
  {\bibfnamefont {M.}~\bibnamefont {Weitz}},\ }\bibfield  {title} {\bibinfo
  {title} {Thermalization kinetics of light: From laser dynamics to equilibrium
  condensation of photons},\ }\href
  {https://doi.org/10.1103/PhysRevA.92.011602} {\bibfield  {journal} {\bibinfo
  {journal} {Phys. Rev. A}\ }\textbf {\bibinfo {volume} {92}},\ \bibinfo
  {pages} {011602} (\bibinfo {year} {2015})}\BibitemShut {NoStop}%
\bibitem [{\citenamefont {Schmitt}\ \emph {et~al.}(2014)\citenamefont
  {Schmitt}, \citenamefont {Damm}, \citenamefont {Dung}, \citenamefont
  {Vewinger}, \citenamefont {Klaers},\ and\ \citenamefont
  {Weitz}}]{Schmitt:2014}%
  \BibitemOpen
  \bibfield  {author} {\bibinfo {author} {\bibfnamefont {J.}~\bibnamefont
  {Schmitt}}, \bibinfo {author} {\bibfnamefont {T.}~\bibnamefont {Damm}},
  \bibinfo {author} {\bibfnamefont {D.}~\bibnamefont {Dung}}, \bibinfo {author}
  {\bibfnamefont {F.}~\bibnamefont {Vewinger}}, \bibinfo {author}
  {\bibfnamefont {J.}~\bibnamefont {Klaers}},\ and\ \bibinfo {author}
  {\bibfnamefont {M.}~\bibnamefont {Weitz}},\ }\bibfield  {title} {\bibinfo
  {title} {Observation of grand-canonical number statistics in a photon
  {B}ose-{E}instein condensate},\ }\href
  {https://doi.org/10.1103/PhysRevLett.112.030401} {\bibfield  {journal}
  {\bibinfo  {journal} {Phys. Rev. Lett.}\ }\textbf {\bibinfo {volume} {112}},\
  \bibinfo {pages} {030401} (\bibinfo {year} {2014})}\BibitemShut {NoStop}%
\bibitem [{\citenamefont {\"Ozt\"urk}\ \emph {et~al.}(2023)\citenamefont
  {\"Ozt\"urk}, \citenamefont {Vewinger}, \citenamefont {Weitz},\ and\
  \citenamefont {Schmitt}}]{Oeztuerk:2023}%
  \BibitemOpen
  \bibfield  {author} {\bibinfo {author} {\bibfnamefont {F.~E.}\ \bibnamefont
  {\"Ozt\"urk}}, \bibinfo {author} {\bibfnamefont {F.}~\bibnamefont
  {Vewinger}}, \bibinfo {author} {\bibfnamefont {M.}~\bibnamefont {Weitz}},\
  and\ \bibinfo {author} {\bibfnamefont {J.}~\bibnamefont {Schmitt}},\
  }\bibfield  {title} {\bibinfo {title} {Fluctuation-dissipation relation for a
  {B}ose-{E}instein condensate of photons},\ }\href
  {https://doi.org/10.1103/PhysRevLett.130.033602} {\bibfield  {journal}
  {\bibinfo  {journal} {Phys. Rev. Lett.}\ }\textbf {\bibinfo {volume} {130}},\
  \bibinfo {pages} {033602} (\bibinfo {year} {2023})}\BibitemShut {NoStop}%
\bibitem [{\citenamefont {Sazhin}\ \emph {et~al.}(2024)\citenamefont {Sazhin},
  \citenamefont {Gladilin}, \citenamefont {Erglis}, \citenamefont {Hellmann},
  \citenamefont {Vewinger}, \citenamefont {Weitz}, \citenamefont {Wouters},\
  and\ \citenamefont {Schmitt}}]{Sazhin:2024}%
  \BibitemOpen
  \bibfield  {author} {\bibinfo {author} {\bibfnamefont {A.}~\bibnamefont
  {Sazhin}}, \bibinfo {author} {\bibfnamefont {V.~N.}\ \bibnamefont
  {Gladilin}}, \bibinfo {author} {\bibfnamefont {A.}~\bibnamefont {Erglis}},
  \bibinfo {author} {\bibfnamefont {G.}~\bibnamefont {Hellmann}}, \bibinfo
  {author} {\bibfnamefont {F.}~\bibnamefont {Vewinger}}, \bibinfo {author}
  {\bibfnamefont {M.}~\bibnamefont {Weitz}}, \bibinfo {author} {\bibfnamefont
  {M.}~\bibnamefont {Wouters}},\ and\ \bibinfo {author} {\bibfnamefont
  {J.}~\bibnamefont {Schmitt}},\ }\bibfield  {title} {\bibinfo {title}
  {{Observation of nonlinear response and Onsager regression in a photon
  Bose-Einstein condensate}},\ }\href
  {https://doi.org/10.1038/s41467-024-49064-9} {\bibfield  {journal} {\bibinfo
  {journal} {Nat. Commun.}\ }\textbf {\bibinfo {volume} {15}},\ \bibinfo
  {pages} {4730} (\bibinfo {year} {2024})}\BibitemShut {NoStop}%
\bibitem [{\citenamefont {Bode}\ \emph {et~al.}(2024)\citenamefont {Bode},
  \citenamefont {Kajan}, \citenamefont {Meirinhos},\ and\ \citenamefont
  {Kroha}}]{Bode:2024}%
  \BibitemOpen
  \bibfield  {author} {\bibinfo {author} {\bibfnamefont {T.}~\bibnamefont
  {Bode}}, \bibinfo {author} {\bibfnamefont {M.}~\bibnamefont {Kajan}},
  \bibinfo {author} {\bibfnamefont {F.}~\bibnamefont {Meirinhos}},\ and\
  \bibinfo {author} {\bibfnamefont {J.}~\bibnamefont {Kroha}},\ }\bibfield
  {title} {\bibinfo {title} {Non-markovian dynamics of open quantum systems via
  auxiliary particles with exact operator constraint},\ }\href
  {https://doi.org/10.1103/PhysRevResearch.6.013220} {\bibfield  {journal}
  {\bibinfo  {journal} {Phys. Rev. Res.}\ }\textbf {\bibinfo {volume} {6}},\
  \bibinfo {pages} {013220} (\bibinfo {year} {2024})}\BibitemShut {NoStop}%
\bibitem [{\citenamefont {Kirton}\ and\ \citenamefont
  {Keeling}(2013)}]{Kirton:2013}%
  \BibitemOpen
  \bibfield  {author} {\bibinfo {author} {\bibfnamefont {P.}~\bibnamefont
  {Kirton}}\ and\ \bibinfo {author} {\bibfnamefont {J.}~\bibnamefont
  {Keeling}},\ }\bibfield  {title} {\bibinfo {title} {Nonequilibrium model of
  photon condensation},\ }\href
  {https://doi.org/10.1103/PhysRevLett.111.100404} {\bibfield  {journal}
  {\bibinfo  {journal} {Phys. Rev. Lett.}\ }\textbf {\bibinfo {volume} {111}},\
  \bibinfo {pages} {100404} (\bibinfo {year} {2013})}\BibitemShut {NoStop}%
\bibitem [{\citenamefont {Hesten}\ \emph {et~al.}(2018)\citenamefont {Hesten},
  \citenamefont {Nyman},\ and\ \citenamefont {Mintert}}]{Hesten:2018}%
  \BibitemOpen
  \bibfield  {author} {\bibinfo {author} {\bibfnamefont {H.~J.}\ \bibnamefont
  {Hesten}}, \bibinfo {author} {\bibfnamefont {R.~A.}\ \bibnamefont {Nyman}},\
  and\ \bibinfo {author} {\bibfnamefont {F.}~\bibnamefont {Mintert}},\
  }\bibfield  {title} {\bibinfo {title} {Decondensation in nonequilibrium
  photonic condensates: When less is more},\ }\href
  {https://doi.org/10.1103/PhysRevLett.120.040601} {\bibfield  {journal}
  {\bibinfo  {journal} {Phys. Rev. Lett.}\ }\textbf {\bibinfo {volume} {120}},\
  \bibinfo {pages} {040601} (\bibinfo {year} {2018})}\BibitemShut {NoStop}%
\bibitem [{\citenamefont {\"Ozt\"urk}\ \emph {et~al.}(2021)\citenamefont
  {\"Ozt\"urk}, \citenamefont {Lappe}, \citenamefont {Hellmann}, \citenamefont
  {Schmitt}, \citenamefont {Klaers}, \citenamefont {Vewinger},\ and\
  \citenamefont {Weitz}}]{Oeztuerk:2021}%
  \BibitemOpen
  \bibfield  {author} {\bibinfo {author} {\bibfnamefont {F.~E.}\ \bibnamefont
  {\"Ozt\"urk}}, \bibinfo {author} {\bibfnamefont {T.}~\bibnamefont {Lappe}},
  \bibinfo {author} {\bibfnamefont {G.}~\bibnamefont {Hellmann}}, \bibinfo
  {author} {\bibfnamefont {J.}~\bibnamefont {Schmitt}}, \bibinfo {author}
  {\bibfnamefont {J.}~\bibnamefont {Klaers}}, \bibinfo {author} {\bibfnamefont
  {F.}~\bibnamefont {Vewinger}},\ and\ \bibinfo {author} {\bibfnamefont
  {M.}~\bibnamefont {Weitz}},\ }\bibfield  {title} {\bibinfo {title}
  {Observation of a non-{H}ermitian phase transition in an optical quantum
  gas},\ }\href {https://doi.org/10.1126/science.abe9869} {\bibfield  {journal}
  {\bibinfo  {journal} {Science}\ }\textbf {\bibinfo {volume} {372}},\ \bibinfo
  {pages} {88} (\bibinfo {year} {2021})}\BibitemShut {NoStop}%
\bibitem [{\citenamefont {Stamper-Kurn}\ \emph {et~al.}(1999)\citenamefont
  {Stamper-Kurn}, \citenamefont {Chikkatur}, \citenamefont {G\"orlitz},
  \citenamefont {Inouye}, \citenamefont {Gupta}, \citenamefont {Pritchard},\
  and\ \citenamefont {Ketterle}}]{Stamper-Kurn:1999b}%
  \BibitemOpen
  \bibfield  {author} {\bibinfo {author} {\bibfnamefont {D.~M.}\ \bibnamefont
  {Stamper-Kurn}}, \bibinfo {author} {\bibfnamefont {A.~P.}\ \bibnamefont
  {Chikkatur}}, \bibinfo {author} {\bibfnamefont {A.}~\bibnamefont
  {G\"orlitz}}, \bibinfo {author} {\bibfnamefont {S.}~\bibnamefont {Inouye}},
  \bibinfo {author} {\bibfnamefont {S.}~\bibnamefont {Gupta}}, \bibinfo
  {author} {\bibfnamefont {D.~E.}\ \bibnamefont {Pritchard}},\ and\ \bibinfo
  {author} {\bibfnamefont {W.}~\bibnamefont {Ketterle}},\ }\bibfield  {title}
  {\bibinfo {title} {Excitation of phonons in a {B}ose-{E}instein condensate by
  light scattering},\ }\href {https://doi.org/10.1103/PhysRevLett.83.2876}
  {\bibfield  {journal} {\bibinfo  {journal} {Phys. Rev. Lett.}\ }\textbf
  {\bibinfo {volume} {83}},\ \bibinfo {pages} {2876} (\bibinfo {year}
  {1999})}\BibitemShut {NoStop}%
\bibitem [{\citenamefont {Utsunomiya}\ \emph {et~al.}(2008)\citenamefont
  {Utsunomiya}, \citenamefont {Tian}, \citenamefont {Roumpos}, \citenamefont
  {Lai}, \citenamefont {Kumada}, \citenamefont {Fujisawa}, \citenamefont
  {Kuwata-Gonokami}, \citenamefont {L{\"o}ffler}, \citenamefont {H{\"o}fling},
  \citenamefont {Forchel},\ and\ \citenamefont {Yamamoto}}]{Utsunomiya:2008}%
  \BibitemOpen
  \bibfield  {author} {\bibinfo {author} {\bibfnamefont {S.}~\bibnamefont
  {Utsunomiya}}, \bibinfo {author} {\bibfnamefont {L.}~\bibnamefont {Tian}},
  \bibinfo {author} {\bibfnamefont {G.}~\bibnamefont {Roumpos}}, \bibinfo
  {author} {\bibfnamefont {C.~W.}\ \bibnamefont {Lai}}, \bibinfo {author}
  {\bibfnamefont {N.}~\bibnamefont {Kumada}}, \bibinfo {author} {\bibfnamefont
  {T.}~\bibnamefont {Fujisawa}}, \bibinfo {author} {\bibfnamefont
  {M.}~\bibnamefont {Kuwata-Gonokami}}, \bibinfo {author} {\bibfnamefont
  {A.}~\bibnamefont {L{\"o}ffler}}, \bibinfo {author} {\bibfnamefont
  {S.}~\bibnamefont {H{\"o}fling}}, \bibinfo {author} {\bibfnamefont
  {A.}~\bibnamefont {Forchel}},\ and\ \bibinfo {author} {\bibfnamefont
  {Y.}~\bibnamefont {Yamamoto}},\ }\bibfield  {title} {\bibinfo {title}
  {Observation of {B}ogoliubov excitations in exciton-polariton condensates},\
  }\href {https://doi.org/10.1038/nphys1034} {\bibfield  {journal} {\bibinfo
  {journal} {Nat. Phys.}\ }\textbf {\bibinfo {volume} {4}},\ \bibinfo {pages}
  {700} (\bibinfo {year} {2008})}\BibitemShut {NoStop}%
\bibitem [{\citenamefont {Schmitt}(2018)}]{Schmitt:2018}%
  \BibitemOpen
  \bibfield  {author} {\bibinfo {author} {\bibfnamefont {J.}~\bibnamefont
  {Schmitt}},\ }\bibfield  {title} {\bibinfo {title} {Dynamics and correlations
  of a {B}ose{\textendash}{E}instein condensate of photons},\ }\href
  {https://doi.org/10.1088/1361-6455/aad409} {\bibfield  {journal} {\bibinfo
  {journal} {J. Phys. B: At. Mol. Opt. Phys.}\ }\textbf {\bibinfo {volume}
  {51}},\ \bibinfo {pages} {173001} (\bibinfo {year} {2018})}\BibitemShut
  {NoStop}%
\bibitem [{sup()}]{supplementalMaterial}%
  \BibitemOpen
  \href@noop {} {}\bibinfo {note} {See {S}upplementary
  {I}nformation.}\BibitemShut {Stop}%
\bibitem [{\citenamefont {Schmitt}\ \emph {et~al.}(2024)\citenamefont
  {Schmitt}, \citenamefont {Weitz},\ and\ \citenamefont
  {Klaers}}]{Schmitt:2024}%
  \BibitemOpen
  \bibfield  {author} {\bibinfo {author} {\bibfnamefont {J.}~\bibnamefont
  {Schmitt}}, \bibinfo {author} {\bibfnamefont {M.}~\bibnamefont {Weitz}},\
  and\ \bibinfo {author} {\bibfnamefont {J.}~\bibnamefont {Klaers}},\
  }\bibfield  {title} {\bibinfo {title} {Absorption and emission spectral data
  of room-temperature rhodamine {6G} dye solution and some typical dye
  microcavity parameters},\ }\href {https://doi.org/10.5281/zenodo.10852936}
  {10.5281/zenodo.10852936} (\bibinfo {year} {2024})\BibitemShut {NoStop}%
\bibitem [{\citenamefont {Buhmann}\ and\ \citenamefont
  {Erglis}(2022)}]{Erglis:2022}%
  \BibitemOpen
  \bibfield  {author} {\bibinfo {author} {\bibfnamefont {S.~Y.}\ \bibnamefont
  {Buhmann}}\ and\ \bibinfo {author} {\bibfnamefont {A.}~\bibnamefont
  {Erglis}},\ }\bibfield  {title} {\bibinfo {title}
  {{Nested-open-quantum-systems approach to photonic Bose-Einstein
  condensation}},\ }\href {https://doi.org/10.1103/PhysRevA.106.063722}
  {\bibfield  {journal} {\bibinfo  {journal} {Phys. Rev. A}\ }\textbf {\bibinfo
  {volume} {106}},\ \bibinfo {pages} {063722} (\bibinfo {year}
  {2022})}\BibitemShut {NoStop}%
\bibitem [{\citenamefont {Rahmani}\ \emph {et~al.}(2024)\citenamefont
  {Rahmani}, \citenamefont {Opala},\ and\ \citenamefont
  {Matuszewski}}]{Rahmani:2024}%
  \BibitemOpen
  \bibfield  {author} {\bibinfo {author} {\bibfnamefont {A.}~\bibnamefont
  {Rahmani}}, \bibinfo {author} {\bibfnamefont {A.}~\bibnamefont {Opala}},\
  and\ \bibinfo {author} {\bibfnamefont {M.}~\bibnamefont {Matuszewski}},\
  }\bibfield  {title} {\bibinfo {title} {Exceptional points and phase
  transitions in non-{H}ermitian nonlinear binary systems},\ }\href
  {https://doi.org/10.1103/PhysRevB.109.085311} {\bibfield  {journal} {\bibinfo
   {journal} {Phys. Rev. B}\ }\textbf {\bibinfo {volume} {109}},\ \bibinfo
  {pages} {085311} (\bibinfo {year} {2024})}\BibitemShut {NoStop}%
\bibitem [{\citenamefont {Chiocchetta}\ \emph {et~al.}(2017)\citenamefont
  {Chiocchetta}, \citenamefont {Gambassi},\ and\ \citenamefont
  {Carusotto}}]{Chiocchetta:2017b}%
  \BibitemOpen
  \bibfield  {author} {\bibinfo {author} {\bibfnamefont {A.}~\bibnamefont
  {Chiocchetta}}, \bibinfo {author} {\bibfnamefont {A.}~\bibnamefont
  {Gambassi}},\ and\ \bibinfo {author} {\bibfnamefont {I.}~\bibnamefont
  {Carusotto}},\ }\bibinfo {title} {Laser operation and {B}ose-{E}instein
  condensation: {A}nalogies and differences},\ in\ \href
  {https://doi.org/10.1017/9781316084366.022} {\emph {\bibinfo {booktitle}
  {Universal {T}hemes of {B}ose-{E}instein {C}ondensation}}},\ \bibinfo
  {editor} {edited by\ \bibinfo {editor} {\bibfnamefont {N.~P.}\ \bibnamefont
  {Proukakis}}, \bibinfo {editor} {\bibfnamefont {D.~W.}\ \bibnamefont
  {Snoke}},\ and\ \bibinfo {editor} {\bibfnamefont {P.~B.}\ \bibnamefont
  {Littlewood}}}\ (\bibinfo  {publisher} {{C}ambridge {U}niversity {P}ress},\
  \bibinfo {year} {2017})\ pp.\ \bibinfo {pages} {409–423,
  \href{https://arxiv.org/abs/1503.02816}{ arXiv:1503.02816}}\BibitemShut
  {NoStop}%
\bibitem [{\citenamefont {Estrecho}\ \emph {et~al.}(2021)\citenamefont
  {Estrecho}, \citenamefont {Pieczarka}, \citenamefont {Wurdack}, \citenamefont
  {Steger}, \citenamefont {West}, \citenamefont {Pfeiffer}, \citenamefont
  {Snoke}, \citenamefont {Truscott},\ and\ \citenamefont
  {Ostrovskaya}}]{Estrecho:2021}%
  \BibitemOpen
  \bibfield  {author} {\bibinfo {author} {\bibfnamefont {E.}~\bibnamefont
  {Estrecho}}, \bibinfo {author} {\bibfnamefont {M.}~\bibnamefont {Pieczarka}},
  \bibinfo {author} {\bibfnamefont {M.}~\bibnamefont {Wurdack}}, \bibinfo
  {author} {\bibfnamefont {M.}~\bibnamefont {Steger}}, \bibinfo {author}
  {\bibfnamefont {K.}~\bibnamefont {West}}, \bibinfo {author} {\bibfnamefont
  {L.~N.}\ \bibnamefont {Pfeiffer}}, \bibinfo {author} {\bibfnamefont {D.~W.}\
  \bibnamefont {Snoke}}, \bibinfo {author} {\bibfnamefont {A.~G.}\ \bibnamefont
  {Truscott}},\ and\ \bibinfo {author} {\bibfnamefont {E.~A.}\ \bibnamefont
  {Ostrovskaya}},\ }\bibfield  {title} {\bibinfo {title} {Low-energy collective
  oscillations and {B}ogoliubov sound in an exciton-polariton condensate},\
  }\href {https://doi.org/10.1103/PhysRevLett.126.075301} {\bibfield  {journal}
  {\bibinfo  {journal} {Phys. Rev. Lett.}\ }\textbf {\bibinfo {volume} {126}},\
  \bibinfo {pages} {075301} (\bibinfo {year} {2021})}\BibitemShut {NoStop}%
\bibitem [{\citenamefont {Busley}\ \emph {et~al.}(2022)\citenamefont {Busley},
  \citenamefont {{Espert Miranda}}, \citenamefont {Redmann}, \citenamefont
  {Kurtscheid}, \citenamefont {{Karkihalli Umesh}}, \citenamefont {Vewinger},
  \citenamefont {Weitz},\ and\ \citenamefont {Schmitt}}]{Busley:2022}%
  \BibitemOpen
  \bibfield  {author} {\bibinfo {author} {\bibfnamefont {E.}~\bibnamefont
  {Busley}}, \bibinfo {author} {\bibfnamefont {L.}~\bibnamefont {{Espert
  Miranda}}}, \bibinfo {author} {\bibfnamefont {A.}~\bibnamefont {Redmann}},
  \bibinfo {author} {\bibfnamefont {C.}~\bibnamefont {Kurtscheid}}, \bibinfo
  {author} {\bibfnamefont {K.}~\bibnamefont {{Karkihalli Umesh}}}, \bibinfo
  {author} {\bibfnamefont {F.}~\bibnamefont {Vewinger}}, \bibinfo {author}
  {\bibfnamefont {M.}~\bibnamefont {Weitz}},\ and\ \bibinfo {author}
  {\bibfnamefont {J.}~\bibnamefont {Schmitt}},\ }\bibfield  {title} {\bibinfo
  {title} {Compressibility and the equation of state of an optical quantum gas
  in a box},\ }\href {https://doi.org/10.1126/science.abm2543} {\bibfield
  {journal} {\bibinfo  {journal} {Science}\ }\textbf {\bibinfo {volume}
  {375}},\ \bibinfo {pages} {1403} (\bibinfo {year} {2022})}\BibitemShut
  {NoStop}%
\bibitem [{\citenamefont {Wanjura}\ \emph {et~al.}(2020)\citenamefont
  {Wanjura}, \citenamefont {Brunelli},\ and\ \citenamefont
  {Nunnenkamp}}]{Wanjura:2020}%
  \BibitemOpen
  \bibfield  {author} {\bibinfo {author} {\bibfnamefont {C.~C.}\ \bibnamefont
  {Wanjura}}, \bibinfo {author} {\bibfnamefont {M.}~\bibnamefont {Brunelli}},\
  and\ \bibinfo {author} {\bibfnamefont {A.}~\bibnamefont {Nunnenkamp}},\
  }\bibfield  {title} {\bibinfo {title} {Topological framework for directional
  amplification in driven-dissipative cavity arrays},\ }\href
  {https://doi.org/10.1038/s41467-020-16863-9} {\bibfield  {journal} {\bibinfo
  {journal} {Nat. Commun.}\ }\textbf {\bibinfo {volume} {11}},\ \bibinfo
  {pages} {3149} (\bibinfo {year} {2020})}\BibitemShut {NoStop}%
\bibitem [{\citenamefont {Garbe}\ \emph {et~al.}(2023)\citenamefont {Garbe},
  \citenamefont {Minoguchi}, \citenamefont {Huber},\ and\ \citenamefont
  {Rabl}}]{Garbe:2024}%
  \BibitemOpen
  \bibfield  {author} {\bibinfo {author} {\bibfnamefont {L.}~\bibnamefont
  {Garbe}}, \bibinfo {author} {\bibfnamefont {Y.}~\bibnamefont {Minoguchi}},
  \bibinfo {author} {\bibfnamefont {J.}~\bibnamefont {Huber}},\ and\ \bibinfo
  {author} {\bibfnamefont {P.}~\bibnamefont {Rabl}},\ }\bibfield  {title}
  {\bibinfo {title} {The bosonic skin effect: boundary condensation in
  asymmetric transport},\ }\href
  {https://doi.org/10.21468/SciPostPhys.16.1.029} {\bibfield  {journal}
  {\bibinfo  {journal} {SciPost Phys.}\ }\textbf {\bibinfo {volume} {16}},\
  \bibinfo {pages} {029} (\bibinfo {year} {2023})}\BibitemShut {NoStop}%
\bibitem [{\citenamefont {Erglis}\ \emph {et~al.}(2025)\citenamefont {Erglis},
  \citenamefont {Sazhin}, \citenamefont {Vewinger}, \citenamefont {Weitz},
  \citenamefont {Buhmann},\ and\ \citenamefont {Schmitt}}]{Erglis:2025data}%
  \BibitemOpen
  \bibfield  {author} {\bibinfo {author} {\bibfnamefont {A.}~\bibnamefont
  {Erglis}}, \bibinfo {author} {\bibfnamefont {A.}~\bibnamefont {Sazhin}},
  \bibinfo {author} {\bibfnamefont {F.}~\bibnamefont {Vewinger}}, \bibinfo
  {author} {\bibfnamefont {M.}~\bibnamefont {Weitz}}, \bibinfo {author}
  {\bibfnamefont {S.~Y.}\ \bibnamefont {Buhmann}},\ and\ \bibinfo {author}
  {\bibfnamefont {J.}~\bibnamefont {Schmitt}},\ }\bibfield  {title} {\bibinfo
  {title} {{Research data supporting "Time-periodic driving of a bath-coupled
  open quantum gas of light"}},\ }\href {10.5281/zenodo.15714719} {\bibfield
  {journal} {\bibinfo  {journal} {Zenodo repository}\ } (\bibinfo {year}
  {2025})}\BibitemShut {NoStop}%
\bibitem [{\citenamefont {Morin}(2008)}]{Morin:2008}%
  \BibitemOpen
  \bibfield  {author} {\bibinfo {author} {\bibfnamefont {D.}~\bibnamefont
  {Morin}},\ }\href@noop {} {\emph {\bibinfo {title} {Introduction to Classical
  Mechanics: With Problems and Solutions}}}\ (\bibinfo  {publisher} {Cambridge
  University Press},\ \bibinfo {year} {2008})\BibitemShut {NoStop}%
\end{thebibliography}

%


\setcounter{figure}{0}
\setcounter{equation}{0}

\newcommand{\ai}{B_\mathrm{abs}}
\newcommand{\ei}{B_\mathrm{em}}

\section*{Supplementary Information}

This Supplementary Material discusses details on the experimental setup, as well as the data acquisition and analysis. Moreover, we include additional figures that illustrate the characteristics of the time-periodic driving scheme. The final section gives a derivation of the photon number response spectrum starting from the coupled rate equation model.

\setcounter{figure}{0}
\renewcommand\thefigure{S\arabic{figure}} 
\renewcommand\theequation{S\arabic{equation}} 

\subsection{Photon gas in dye-filled bispherical optical microcavity}

We prepare the photon Bose--Einstein condensate (BEC) in an optical microcavity (see the left panel of Fig.~1(a) of the main text) consisting of two spherically-curved mirrors with a reflectivity $>99.997\%$. To allow for a photon thermalization, we fill the cavity volume with a liquid dye solution (Rhodamine 6G molecules with a concentration of $\SI{1}{\milli\mol\per\liter}$ dissolved in ethylene glycol) with the refractive index of $n_{\mathrm{r}}\approx 1.44$. Through multiple absorption and emission cycles with the molecules, photons acquire a Bose--Einstein distribution with the temperature of the dye, $T=\SI{300}{\kelvin}$. The short cavity length $D_0=\SI{1.4}{\micro\meter}$ determines the free spectral range $\Delta \nu_\mathrm{FSR}=c/(2 n_{\mathrm{r}}D_0) \approx \SI{74}{\tera \hertz}$, which is much larger than the spectral width of the dye molecules determined by thermal energy $\kB T/\hbar \approx 2\pi\times\SI{6.2}{\tera\hertz}$. The energy gap freezes out the longitudinal degree of freedom of the photons in the cavity, making the photon gas effectively two-dimensional. Moreover, the cavity length establishes a minimal 'cutoff' frequency of the photons, $\omega_{\mathrm{c}} = \pi c q/(n_{\mathrm{r}} D_0)  \approx 2\pi\times \SI{520}{\tera\hertz}$ with longitudinal wavenumber $q$ discussed in the main text, which ensures that the creation of photons in the cavity by thermal excitations is strongly suppressed by $\exp(-\hbar\omega_{\mathrm{c}}/\kB T)\sim 10^{-36}$. The decoupling of photon number and temperature allows us to independently control both quantities. This results in a nonvanishing chemical potential of the photon gas $\mu\neq 0$, in contrast to a blackbody radiator. Furthermore, it guarantees that the photon number is conserved throughout the thermalization process. The cavity geometry, which is determined by the mirror curvature, imposes transverse confining potential for the photon gas. For spherically curved mirrors, it can be shown, that the photon gas is confined in a harmonic trapping potential~\cite{Schmitt:2018}. The trapping frequency is calculated as $\omega_{\mathrm{trap}}=c/n_\mathrm{r} \sqrt{2/(D_0 R)} \approx 2\pi\times \SI{40}{\giga\hertz}$, where $R=\SI{1}{\meter}$ is the radius of curvature of the mirrors. Overall, this yields the energy-momentum relation of the two-dimensional photon gas in the bispherical microcavity
\begin{equation}
    E (k_r)= m_\mathrm{ph} \frac{c^2}{n_{\mathrm{r}}^2}+\frac{\hbar^2 k_r^2}{2 m_\mathrm{ph}} + \frac12 m_\mathrm{ph} \omega_{\mathrm{trap}}^2 r^2.
    \label{eq:dispersion}
\end{equation}
Here we have introduced an effective photon mass $m_\mathrm{ph}=\pi\hbar q n_{\mathrm{r}}/(D_0 c)\approx \SI{8e-36}{\kilo\gram}$, which is given by the cavity cutoff frequency $\omega_\mathrm{c}$. A detailed derivation of Eq.~\eqref{eq:dispersion} can be found in Ref.~\cite{Schmitt:2018}. The transverse eigenenergies in the cavity read
\begin{equation}
    \epsilon_{u,v}=\hbar \omega_{\mathrm{trap}}(u+v),
\end{equation}  
with quantum numbers $(u,v)=(0,0),(1,0),(0,1)$ etc. The energy levels exhibit a degeneracy $g=2(u+v+1)$, where the prefactor accounts for the two-fold polarization degeneracy of the modes. The average total photon number $\bar{N}$ in the cavity is calculated as a sum of the Bose--Einstein distribution over energies $\epsilon_{u,v}$,
\begin{equation}
    \bar{N}=\sum_{u,v}\ddfrac{g(\epsilon_{u,v})}{\exp[(\epsilon_{u,v}-\mu)/\kB T]-1}.
    \label{eq:BECsum}
\end{equation}
The measured photon number distribution is shown as a function of the wavelength in Fig.~1(a) (right panel) of the main text. The eigenmodes are given by Hermite--Gaussian modes in the transverse $x$ and $y$ directions
\begin{align}
    \psi_{u,v}(x,y)= & \frac{1}{\pi b^2\sqrt{2^{u+v}u!v!}} \nonumber \\
    & \times H_u\left(\frac{x}{b}\right) H_v\left(\frac{y}{b}\right) \exp(-\frac{x^2+y^2}{2b^2}),
\end{align}
where $b=\sqrt{\hbar/m_{\mathrm{ph}}\omega_{\mathrm{trap}}}$ and
$H_u(x)$ is the $u$-th order Hermite polynomial. The condensate populates the TEM$_{00}$ mode with a probability density $|\psi_{0,0}(x,y)|^2$, which is the state with the lowest available energy in the system. The condensate mode has a FWHM diameter of $d=2\sqrt{\hbar \ln 2/m_{\mathrm{ph}}\omega_{\mathrm{trap}}} \approx \SI{12.2}{\micro\meter}$. The phase transition of the thermalized photon gas to the BEC occurs at the critical photon number
\begin{equation}
    \Nc=\frac{\pi^2}{3}\left(\frac{\kB T}{\hbar \omega_{\mathrm{trap}}}\right)^2.
\end{equation}
This expression is obtained by summing the population in the excited states in Eq.~\eqref{eq:BECsum} for $\mu=0$. For $\bar{N}<N_{\mathrm{c}}$, all particles occupy only the excited cavity modes. At $\bar{N}=N_{\mathrm{c}}$, the population in the excited states starts to saturate; for $\bar{N}>\Nc$, all photons which are added to the system in excess of $\Nc$ occupy the ground state of the system and the mean condensate occupancy is $\nb=\bar{N}-\Nc$.

In the experiment, we inject the photon gas into the cavity by nonresonantly pumping the dye medium with a laser beam. To probe the frequency response of the condensate mode, the pump laser power is temporally modulated. To avoid spatial inhomogeneities in the process of periodically-driving the system, the laser pump beam size is chosen to have a diameter of roughly $\SI{100}{\micro\meter}$, which is larger than the spatial extent of the condensate. In this way, we ensure that all the molecules interacting with the condensate mode are pumped homogeneously. In particular, this prerequisite allows us to model the coupling between the condensate and the dye molecules by position-independent rate equations, see Eq.~(1) of the main text.

\subsection{Time-periodic driving}

\begin{figure*}
    \centering
    \includegraphics[width=0.7\textwidth]{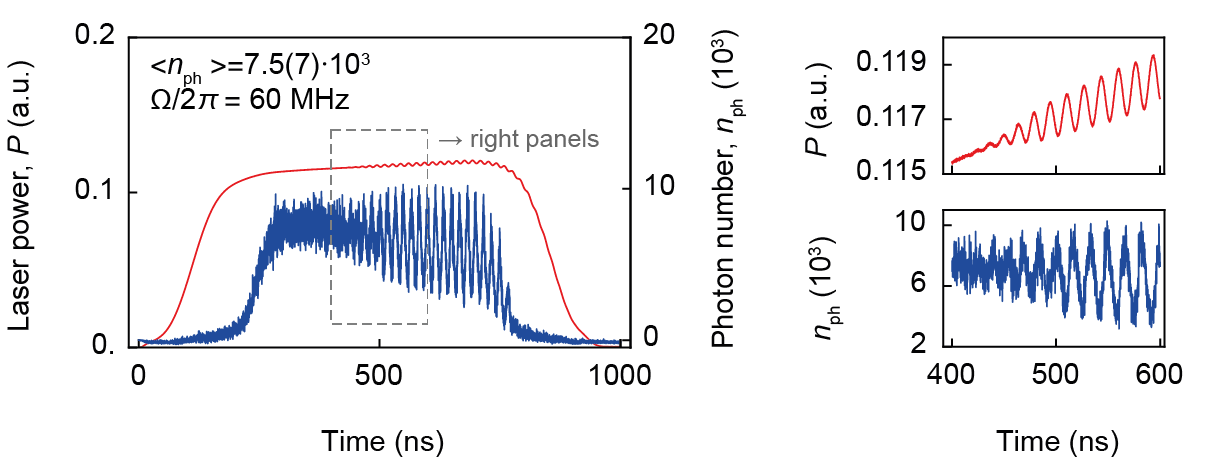}
    \caption{Pump power (red) and photon number (blue) averaged over 100 individual time traces. Here, the average photon number $\nb=7.5\times10^3$ and driving frequency $\Omega/2\pi=\SI{60}{\mega\hertz}$.}
    \label{fig:whole_pulse}
\end{figure*}

To generate a pump beam with a temporally-varying power, three acousto-optic modulators (AOMs) are employed. Initially, two AOMs in series chop the laser light into pulses, producing a flat-top power profile. Next, the beam is directed through a double-pass configuration with a third AOM, shifting the laser frequency. The frequency-shifted beam is then combined with the original frequency beam within a Mach--Zehnder interferometer, enabling tunable oscillations in pump power during the second part of the pulse. The setup, incorporating both the double-pass and Mach--Zehnder interferometer, is outlined in Fig.~1(b) of the main text. A typical time trace of both the pump pulse and the condensate occupancy is presented in Fig.~\ref{fig:whole_pulse}. Further details about the experimental configuration are discussed below.

\vspace{0.5cm}
{\itshape{Pumping scheme.---}} First, we describe our method to optically inject the photon BEC with a well controlled average condensate occupancy by chopping the pump beam into temporal pulses. To pump the molecules in the cavity, we use a continuous-wave laser emitting at $\SI{532}{\nano\meter}$ wavelength (Coherent Verdi 15G). To minimize optical pumping of the dye molecules into the long-lived triplet states, the laser power is segmented into temporal pulses lasting roughly $\SI{600}{\nano\second}$, set at a repetition rate of $\SI{50}{\hertz}$ (see Fig.~\ref{fig:whole_pulse}). For this purpose, two consecutive AOMs are used, both driven with a radiofrequency of $\SI{110}{\mega\hertz}$. The two AOMs ensure sufficient suppression of the pump light reaching the cavity during the 'off' intervals. Furthermore, to maintain an approximately constant average photon number $\nb$ throughout the pump pulse (in the absence of a periodic power modulation, discussed below), the laser power is linearly increased during the pulse, as indicated by the pump laser time trace (red line) in Fig.~\ref{fig:whole_pulse}.

\vspace{0.5cm}
{\itshape{Double-pass interferometric setup: Creating a periodic drive.---}} Next, we turn to the time-periodic modulation of the pump power, which we refer to as 'driving'. To achieve a sinusoidally oscillating laser power, we use a double-pass setup with an additional, third AOM, see Fig.~1(b) of the main text for a scheme of the setup. For this, the output of the above described laser beam is guided through a $\lambda/2$-plate, which rotates the polarization of the beam and allows us to determine the fraction of power being transmitted and reflected in the polarizing beamsplitter (PBS), which ultimately allows us to control the modulation depth of the pumping. The reflected part of the beam passes through another $\lambda/2$-plate, in order to optimize the diffraction efficiency of the AOM. The AOM is driven with variable radiofrequencies of $\Omega_\mathrm{RF}=\Omega/2=2\pi\times\SI{30}{\mega\hertz}$ to $\SI{160}{\mega\hertz}$. The first-order diffracted beam with frequency $\omega_L+\Omega/2$ ($\omega_L$ is the laser frequency) is filtered by an iris, directed through a $\lambda/4$-plate and reflected back. After the second pass of the $\lambda/4$, the linear polarization of beam is rotated by $90^\circ$, and the beam is diffracted once more in first order shifting the frequency to $\omega_L+\Omega$. The beam is transmitted through the PBS into the Mach--Zehnder interferometer. At the nonpolarizing output beamsplitter, the initial (unshifted) and double-pass beam (frequency-shifted) are overlapped and interfere to give an oscillating power.

\vspace{0.5cm}
{\itshape{Time traces of driving and condensate response.---}} In the experiment, we switch on the radiofrequency of the third AOM approximately at the middle point in time of the pulse, as seen in the time trace in Fig.~\ref{fig:whole_pulse} at around $\SI{400}{\nano\second}$. Subsequently, the modulated beam is directed towards the cavity where it pumps the dye medium. The photon BEC partially leaks out of the cavity and is monitored on a fast photomultiplier with a temporal resolution of $\SI{150}{\pico\second}$ (Photek PMT210)~\cite{Sazhin:2024}. The output voltage was sampled on an oscilloscope with $\SI{6}{\giga\hertz}$ bandwidth (Tektronix MSO64B). A typical measured photon number evolution averaged over 100 time traces is shown in Fig.~\ref{fig:whole_pulse} (blue data). Note that these time traces depict the measurement signal dominated by photons occupying the ground mode. For this, a momentum filter was used in the experimental setup to filter out contributions from photons in excited modes that exhibit a larger transverse momentum. In every experimental realization (i.e., every $\SI{600}{\nano\second}$ pulse) the phase of the driving varies with respect to the overall pulse timing. This means that when averaging both the driving laser and the photon condensate response raw data, the oscillation is washed out and not visible. Therefore, each pump pulse was analyzed individually and the phase of the driving laser was extracted to serve as a reference phase $\phi_0$ for both the laser and the condensate trace. For an appropriate averaging and to extract the relative phase between the laser and the condensate, we shifted the time traces by the corresponding reference phases for each experimental realization and only subsequently averaged the data. In the averaged results, shown in Fig.~\ref{fig:whole_pulse}, a clear oscillation is visible indicating that the applied phase correction method works correctly.

The right plots of Fig.~\ref{fig:whole_pulse} give a magnified view of the driving power and the condensate response time traces. Here one can see that due to a finite rise time of the AOM, the driving amplitude does not start instantaneously but takes approximately 5 oscillation periods to reach a stationary oscillation amplitude. The same holds true for the condensate response. Therefore, for all further analysis of the response spectra we only fit the data in the temporal region where the oscillation amplitude has reached a stationary value. Furthermore, the plots highlight that the modulation depth of the drive of around 0.7\% is quite small, while the condensate response depth of $\delta n/\nb\approx 22$\% is substantial, where $\delta n$ denotes the fitted photon number response amplitude. At first glance, the latter may seem a too large value to describe the photon BEC response by linearized equations, i.e., assuming terms of type $\nt\Met$ (see Eq.~(2) of the main text) to be negligible. Experimentally, however, we find that the sinusoidal fits to the data (see Fig.~1(c) of the main text) still describe the oscillations very well even at these modulation depths, which indicates the condensate dynamics is not yet in the regime of nonlinear response. To cross validate our observations, we have also performed numerical calculations of the coupled BEC-bath system, which confirmed that the situation is well in the regime of linear response and that anharmonic contributions to the response spectrum $A(\omega)$ (see Figs.~2 and 3 in the main text) can be safely neglected.

Finally, we comment on how the system parameters cavity loss rate $\kappa$, molecule number $M$, and excited molecule number $\Me$ were determined in the present work. We extracted the total molecule number $M=4.4(3)\times 10^9$ and the cavity loss rate $\kappa=\SI{7.4\pm 0.1}{\giga\hertz}$ from fitting the theoretical predictions for $\Omega_0(\bar n)$ and $\Gamma(\bar n)$, see Eqs.(3) and (4) of the main text, to the two data sets shown in Fig.~3(b) (top, middle panels) of the main text. From this, one obtains the mean excited molecule number
\begin{equation}
    \bar M_e=\frac{B_\mathrm{abs} M  +  \kappa}{\frac{B_\mathrm{em}}{\bar n} + B_\mathrm{abs} + B_\mathrm{em}},
\end{equation}
which with the absorption and emission coefficients $B_\mathrm{abs}=\SI{492}{\hertz}$ and $B_\mathrm{em}=\SI{23.3}{\kilo\hertz}$, respectively, yields $\bar M_e\approx 9.1(6)\times 10^7$. At the investigated photon numbers (of several thousands), this number is approximately independent of $\bar n$.

\begin{figure}
    \centering
\includegraphics[width=0.9\columnwidth]{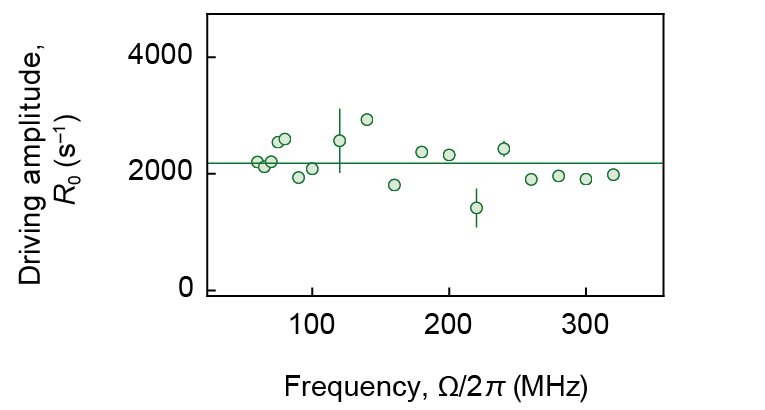}
    \caption{Calibrated driving amplitude versus frequency.}
    \label{fig:R0_dependence}
\end{figure}

\vspace{0.5cm}
{\itshape{Frequency-dependence of the driving amplitude.---}} The response spectra of the photon BEC (see Figs.~2(a), 3(a) of the main text) are extracted from the fitted amplitude of the time traces of the photon number, obtained when driving the dye molecules in a time-periodic fashion with a specific drive strength. For a faithful extraction of the response spectrum, it is therefore important to take into account variations in the driving amplitude. 

Experimentally, variations in the driving amplitude at different driving frequencies $\Omega$ are caused by the laser beam that passes the double-pass setup. First, the diffraction efficiency of the AOM is frequency-dependent, which results in a varying power of the frequency-shifted beam. Secondly, for each driving frequency, the diffraction angle of the first order is slightly modified which modified the amplitude of the beating at the interferometer output. To correct for both experimental imperfections and to ensure an approximately constant driving amplitude, we readjusted the experimental setup for each used driving frequency $\Omega$ and adjusted the used radiofrequency power supplied to the AOM. Despite these corrections, residual variations of roughly $15\%$ (rms) in the driving amplitude as a function of $\Omega$ remained, as shown in Fig.~\ref{fig:R0_dependence}. To take them into account, we normalized the measured response amplitude of the condensate against the respective driving amplitude for each driving frequency.

\widetext
\subsection{Derivation of response spectrum from the rate equation model}\label{sec:PBECHarmOsc}
In this section, we derive the equation of motion for the photon BEC under weak time-periodic driving given in Eq.~(2) of the main text, starting from the photon and excited molecule number rate equations. Its solution yields the response spectrum, the free oscillation frequency $\Omega_0$, and the damping rate $\Gamma$, see Eqs.~(3)--(5) of the main text. At the end, we will see that the spectrally-resolved elementary excitations of the open BEC-bath system are well described by a driven--damped harmonic oscillator.

We start with the rate equations for the photon and excited molecule number, see Eq.~(1) of the main text:
\begin{subequations}\label{eq:rateEquationsExp}
\begin{align}
\label{eq:rateEquationsExp_a}
    \dot{n}=& -\ai \Mg n + \ei \Me (n+1)- \kappa n, \\
\label{eq:rateEquationsExp_b}
    \dot{\Me}=&\ \ai \Mg n  -\ei  \Me (n+1) -\GDown \Me +\left[\GUp + R_0 \sin(\Omega t) \right]\Mg.
\end{align}\end{subequations}
Here, $n$ is the photon number in the condensate, $\Mg$ and $\Me$ are the numbers of molecules in the ground and excited state, respectively, $\GDown$ is the spontaneous decay, $\GUp$ ($R_0$) is the steady (oscillating) pumping rate provided by the laser beam, and $\Omega$ is the driving frequency. In comparison to Eq.~(1) of the main text, in Eq.~\eqref{eq:rateEquationsExp_b} we have already inserted the time-modulated pump rate $R(t) = \GUp + R_0\sin(\Omega t)$. The terms $\GUp$ and $R_0$ can be obtained from the semiclassical description of the interaction between the laser beam and a molecule~\cite{Erglis:2022},
\begin{align}
    \GUp & =\frac{\pi d^2_{01}I_0}{c\varepsilon_0\hbar^2},\\
    R_0\sin(\Omega t) & =\frac{\pi d^2_{01}I_1}{c\varepsilon_0\hbar^2} \sin(\Omega t).
\end{align}
Here, $I_0\equiv I_0(\omega_{\mathrm{zpl}})$ and $I_1\equiv I_1(\omega_{\mathrm{zpl}})$ denote the DC and AC part of the laser intensity per unit frequency (with dimension $J \times m^{-2}$) at the frequency of the zero-phonon line $\omega_{\mathrm{zpl}}$, which determines the electronic excitation energy of the molecule, $d_{01}$ the electronic transition dipole moment of the molecule, $\varepsilon_0$ the vacuum permittivity, $c$ the speed of light, and $\hbar$ the Planck constant.

Since our experiment measures photon numbers, we eliminate the molecule dynamics in the limit of small driving amplitudes and obtain a self-consistent equation for the photon number. For this purpose, we start with an ansatz for the photon and molecule numbers:
\begin{subequations}
\begin{align}
    n(t)&=\nb+\abs{n_0(\Omega)}\sin(\Omega t+\phi)  
    =\nb+\nt,\label{eq:nt_sol_osc} \\
    \Me(t)&=\Meb+\abs{M_{\mathrm{e}_0}(\Omega)}\sin(\Omega t+\phi_{\Me})
    =\Meb+\Met, \\
    \Mg(t)&=\Mgb + \abs{M_{\mathrm{g}_0}(\Omega)}\sin(\Omega t+\phi_{\Mg}) =\Mgb+\Mgt. 
\end{align}
\end{subequations}
The steady-state part denoted by bar (--) comes from the time-independent pumping term $\GUp$, while the oscillatory part, denoted by tilde ($\sim$), follows from the time-modulated driving term $R_0\sin(\Omega t)$. Here, $|n_0(\Omega)|$, $|M_{\mathrm{e}_0}(\Omega)|$, and $|M_{\mathrm{g}_0}(\Omega)|$ are the response amplitudes for $n(t)$, $\Me(t)$, and $\Mg(t)$, respectively. Because of the finite response time of the coupled system of photons and molecules, we expect that the solutions in general exhibit different phases, $\phi$, $\phi_{\Me}$, and $\phi_{\Mg}$. Moreover, we have
\begin{align}
    \Mg(t)=M-\Me(t)
    =M-\Meb-\abs{M_{\mathrm{e}_0}(\Omega)}\sin(\Omega t+\phi_{\Me}),
\end{align}
under the assumption that total molecule number $M$ is constant. This is well fulfilled in the experiment under conditions of pulsed pumping with which we ensure to minimize any excitation of molecules to triplet states as well as bleaching effects. Substituting $n(t)=\nb+\nt$ and $\Me(t)=\Meb+\Met$ in Eqs.~\eqref{eq:rateEquationsExp}, one obtains:
\begin{subequations}\label{eq:rateEquationsExp2}
\begin{align}
&\dot{\nb}+\dot{\nt} =  - \ai(\Mgb+\Mgt)(\nb+\nt) + 
     \ei(\Meb+\Met)(\nb+\nt+1)
     -\kappa (\nb+\nt),\\
&\dot{\Meb} + \dot{\Met} =   \ai(\Mgb + \Mgt ) (\nb + \nt)  
    - \ei(\Meb +\Met) (\nb \!+\! \nt + 1) -  \GDown ( \Meb \!+\!\Met)   + \left[ \GUp + R_0 \sin(\Omega t)\right] (\Mgb + \Mgt ).
\end{align}
\end{subequations}
We can simplify the above equations by recognizing that the steady-state solutions are independent of time,
\begin{subequations}\label{eq:steadyEqs}
\begin{align}
    \dot{\nb}=0  = & -\ai\Mgb\nb  + \ei\Meb(\nb+1) - \kappa \nb,\label{eq:steadyEqs1}\\
    \dot{\Meb}=0 =& \  \ai\Mgb\nb  - \ei\Meb(\nb+1) - \GDown \Meb  
     + \GUp\Mgb.
\end{align}
\end{subequations}
Furthermore, in the limit of weak driving we can neglect higher-order nonlinear terms $\nt\Met$ and $\nt\Mgt$. Additionally, we drop the $\GUp \Mgt$ and $R_0\sin(\Omega t)\Mgt$ terms, because due to $\Mgt\approx 10^4 \ll \Mgb\approx 10^9$, $R_0= \mathcal{O}(\SI{1}{\kilo\hertz})$, and $\GUp=\mathcal{O}(\SI{100}{\kilo\hertz})$ these rates are significantly smaller than those occurring in the other terms. With these simplifications, one obtains
\begin{subequations}\label{eq:rateEquationsOsc}
\begin{align}
    \dot{\nt}&=(\ei\Meb-\ai\Mgb-\kappa)\nt+
    [\ei(\nb+1)+\ai\nb]\Met,\label{eq:rateEquationsOsc1}
    \\
    \dot{\Met}&=(\ai\Mgb-\ei\Meb)\nt-\big[\ei(\nb+1)
    +\ai\nb+\GDown\big]\Met+R_0\sin(\Omega t)\Mgb.\label{eq:rateEquationsOsc2}
\end{align}
\end{subequations}
Taking the temporal derivative of Eq.~\eqref{eq:rateEquationsOsc1}, inserting Eq.~\eqref{eq:rateEquationsOsc2} to replace $\dot{\Met}$, and using Eq.~\eqref{eq:rateEquationsOsc1} to express $\Met$, after some simplifications we obtain the equation of motion for the photon number $\nt$:
\begin{equation}\label{eq:ddotn0final}
    \ddot{\nt}+\Gamma \dot{\nt} +\Omega_0^2 \nt = F_0 \sin{\Omega t}.
\end{equation}
This is Eq.~(2) of the main text. The free oscillation frequency $\Omega_0$ and damping rate $\Gamma$ are given by
\begin{align}
    \Omega_0^2&= [\ei(\nb+1)+\ai\nb] \kappa 
    + \GDown(\ai\Mgb-\ei\Meb+\kappa),\label{eq:w0freq2}\\
    \Gamma& =\kappa+(\ai\Mgb-\ei\Meb)
   +[\ei(\nb+1)+\ai\nb]+\GDown. \label{eq:GammaDamp2}
\end{align}

To recast $\Mgb$ in terms of $\Meb$, we use Eq.~\eqref{eq:steadyEqs1}:
\begin{equation}
    \ai\Mgb=\ei\left(1+\frac{1}{\nb}\right)\Meb-\kappa.
\end{equation}
The final expressions for the oscillation frequency and damping now read
\begin{align}
    \Omega_0^2 &= [\ei(\nb+1)+\ai\nb]\kappa+\frac{\ei \Meb}{\nb}\GDown,\label{eq:w0freqFinal}\\
    \Gamma &= [\ei(\nb+1)+\ai\nb]+\frac{\ei \Meb}{\nb}+\GDown. \label{eq:GammaDampFinal}
\end{align}
These are Eqs.~(3) and~(4) of the main text. Equation~\eqref{eq:ddotn0final} can be solved exactly~\cite{Morin:2008}, and one finds the spectrum of the photon number response  
\begin{equation}\label{eq:responseBEC}
    \abs{n_0(\Omega)}=\frac{F_0}{\sqrt{(\Omega^2-\Omega_0^2)^2+\Gamma^2\Omega^2}}.
\end{equation}
This is Eq.~(5) of the main text. The spectrum can be understood as an elementary excitation of the optical quantum gas, which emerges only in the open BEC-bath system due to pumping and loss. This elementary excitation therefore differs in its nature from those observed in polariton or cold atomic gases, which are dominated by interactions~\cite{Utsunomiya:2008,Stamper-Kurn:1999b}. We also obtain the phase delay of the photons number evolution with respect to the driving laser 
\begin{equation}\label{eq:phase_delay2}
    \phi(\Omega)=-\arccos{\frac{\Omega_0^2-\Omega^2}{\sqrt{(\Omega^2-\Omega_0^2)^2+\Gamma^2\Omega^2}}},
\end{equation}
and the imaginary and real parts of the photon BEC response
\begin{align}\label{eq:responseBECIm}
    \Im n_0(\Omega)&=-\abs{n_0(\Omega)}\sin \phi,\\
    \Re n_0(\Omega)&=\abs{n_0(\Omega)}\cos\phi.
\end{align}

\end{document}